\documentclass[12pt,preprint]{aastex}
\slugcomment{Accepted  for publication  
in {\it the Astrophysical Journal}}  
\def\lax {\ifmmode{_<\atop^{\sim}}\else{${_<\atop^{\sim}}$}\fi}  
\def\gax {\ifmmode{_>\atop^{\sim}}\else{${_>\atop^{\sim}}$}\fi}  
\def\gtorder{\mathrel{\raise.3ex\hbox{$>$}\mkern-14mu
             \lower0.6ex\hbox{$\sim$}}}
\def\etal { et al. }

\begin{document}

\title{Comprehensive Analysis of  {\it RXTE}  Data from Cyg X-1:
Spectral Index-Quasi-Periodic Oscillation Frequency-Luminosity  Correlations}

\author{Nickolai Shaposhnikov\altaffilmark{1} and Lev Titarchuk\altaffilmark{2, 3} }

\altaffiltext{1}{Goddard Space Flight Center, NASA, Exploration of the Universe Division/Universities Space Research Association, code 662, Greenbelt  
MD 20771; nikolai@milkyway.gsfc.nasa.gov}
\altaffiltext{2}{Goddard Space Flight Center, NASA, Exploration of the Universe Division, code 661, Greenbelt  
MD 20771; lev@milkyway.gsfc.nasa.gov}
\altaffiltext{3}{George Mason University/Center for Earth
Observing and Space Research, Fairfax, VA 22030; and US Naval Research
Laboratory, Code 7655, Washington, DC 20375-5352; ltitarchuk@ssd5.nrl.navy.mil }

\begin{abstract}
We present timing and spectral analysis of  $\sim$ 2.2 Ms of  Rossi X-ray Time Explorer ({\it RXTE})
 archival data from Cyg X-1. Using a generic Comptonization model we reveal that the spectrum of 
 Cyg X-1 consists of three components: a thermal seed photon spectrum, a Comptonized part of the
  seed photon spectrum and the iron line. We find  a strong correlation between the 0.1-20 Hz frequencies
of quasiperiodic oscillations (QPOs) and the spectral index. Presence of two spectral phases (states) 
are clearly seen in the data when the spectral indices saturate at low and high values 
of QPO frequencies. This saturation effect was  discovered earlier in a number  of black hole  candidate
 (BHC)  sources and now we strongly confirm  this phenomenon  in Cyg X-1.
In the soft state this index-QPO frequency correlation shows a saturation  of the photon index  
$\Gamma\sim 2.1$ at high values of the low frequency $\nu_{L}$. 
The saturation level of $\Gamma\sim 2.1$ is the lowest  value found yet in  BHCs.
The  bolometric luminosity  does not show clear correlation with the index.
We also show that  Fe K$_{\alpha}$ emission line strength (equivalent width, EW) 
 correlates with  the QPO frequency. The EW increases from 200 eV
in the low/hard state to 1.5 keV in the high/soft state.
The observational correlations revealed compel us to propose a scenario for 
the spectral transition and iron line formation which occur in  BHC sources.     
We also present the spectral state (power-law index) evolution for eight 
years of Cyg X-1 observations by  {\it RXTE}. 
\end{abstract}

\keywords{accretion, accretion disks---black hole physics---stars:individual (Cyg X-1)
:radiation mechanisms: nonthermal---physical data and processes}

\section{Introduction}
Cyg X-1 is one of the brightest high-energy sources in the sky, with an average 1-200 keV energy flux 
of $\sim3\times10^{-8}$ ergs cm$^{-2}$s$^{-1}$. Its optical companion is an O9.7 Iab supergiant HDE
 226868. Estimates of the mass, M, of the X-ray star, $5\lax M_{\odot}\lax15$ [e.g., \citet{her95}] strongly 
 suggest the presence of a black hole. Observed spectral and temporal X-ray characteristics are
 extensively studied based on the large amount of  data  collected in the {\it RXTE} archive (see \S 2 for
  the data description). Our analysis includes $\sim$ 2.2 Ms of  Rossi X-ray Time Explorer ({\it RXTE}) 
  archival data from Cyg X-1  to study the spectral and timing properties of this 
  classical BHC source.

One of the  basic questions addressed in many observational and theoretical 
studies concerning relativistic compact objects is how to observationally distinguish between a neutron star (NS) and a black hole (BH). 
Cyg X-1,  being extensively studied, has often been used as the prototypical example of a BH. 
The different  patterns, for example correlations 
between spectral and timing characteristics of BH and NS sources, has been proposed as
a criteria for determination of the nature of the compact object. Recently Belloni (2005) and \citet{mr}
 published a concise review of the observational features of the spectral states in BH sources where 
 they also point out a link between timing and spectral properties of X-ray radiation and plasma
  ejection  leading  to radio jets. 
   
 \citet{tf04}, hereafter TF04, and \citet{tsha05}, hereafter TSh05, present theoretical and observational arguments how to distinguish between BH and NS binaries.
TF04 present observational evidence that in BHs two  distinct 
phases occur: one of them, the steep power-law phase (so ``called'' high/soft state), 
is the signature of a BH.  In the soft state of BH the spectral index-quasiperiodic oscillation 
(QPO) frequency correlation shows a flattening, or  ``saturation'' of the photon
index $ \Gamma$ at high values of the low frequency $\nu_L$.  This saturation effect was identified as a BH signature. 
TSh05 demonstrate that this saturation is not present  in at least one NS source. 
They show that for  4U 1728-34  the index $\Gamma$ increases  monotonically  with $\nu_L$ .
We show here that Cyg X-1 is a perfect example of a BH source as the suggested BH  
index-QPO frequency correlation is  observed with  clear features of the saturation at high  and low
 frequencies. 

Long-term monitoring of Cyg X-1 has revealed two distinct spectral states and transitions between them 
[see a review of Cyg X-1 early observations in
\citet{zhang97}]. Simultaneous, low- and high energy X-ray observations during interstate  
transitions have been
obtained by several groups in 1996 [see references in \citet{zhang97} and \citet{cui97, cui98}]. 
In  the low/hard state,  the power-law portion of the
spectrum is relatively flat  with a photon index $\Gamma$ of  about 1.5. They found that 
a majority of the time Cyg X-1 stayed in 
the low/hard with an occasional transition (during their observations,  duty cycle 
of this state was about 90 \%) 
to the soft state where the power-law spectrum became significantly steeper 
(with $\Gamma\sim 2.5$).  Also, one or two times per year   Cyg X-1 exhibited so called 
"failed state transitons",  when it started to transition but did not reach a soft state, stopping at some
intermediate state  and their falling  back. Thus one can claim that the source was predominantly in a hard state in 90s. 
However  Wilms et al. (2005) have recently shown that since 2000 the source spent $\sim 34\%$ of the time in the intermediate and the soft state. 
We came to the same conclusion as a result of our analysis of the source spectral transition 
 (see more discussion of this phenomenon in \S 4, the last paragraph).

As \citet{cui97} pointed out there is strong evidence that the observed QPO characteristics are related to spectral
properties: the QPO amplitude increases as the energy spectrum becomes harder. They also discovered QPO low-frequencies 
varying in the range of 4-10 Hz during the spectral transition.  
In fact, using these observations Di Matteo \& Psaltis (1999), hereafter DP99, found that the photon index can be correlated with the QPO frequency.
One can see a few points of this correlation in the frequency range from 1 to 10 Hz in their Figure 1. 
This behavior was later confirmed by \citet{pott03}, hereafter P03, using
observations of the Cyg X-1 spectral transition in 2000-2001.
It is worth noting that  DP99 also suggested that 
the index-qpo frequency correlation can be a common phenomenon for  black hole sources.

In this Paper we  present a detailed study of spectral transitions in Cyg X-1 and 
 demonstrate how the energy spectra are related to the power density spectra (PDSs),
in particular the QPO features. 
We find  that  the index-QPO correlation is similar to previous findings for BH 
sources, e.g. DP99,  \citet{vig} and TF04, where  the QPO frequency-index correlation is presented  
for large samples of BH sources.   In  PDSs observed by {\it RXTE} for Cyg X-1,  
we show that these QPO low frequencies tightly  correlate with the break frequency $\nu_b$. 

\citet{to}  presented a model for the radial oscillations and diffusion in the transition layer (TL)
 surrounding the  BH and NS.  Using dimensional analysis, they  identified the corresponding radial 
 oscillation  and diffusion frequencies in the TL with the low-Lorentzian $\nu_{L}$ and break frequencies 
 $\nu_b$  for 4U 1728-34.  They predicted values for $\nu_b$ related to the diffusion in the transition 
 layer, that are consistent with the observed $\nu_b$. Both the Keplerian and radial oscillations, 
along with diffusion in the transition layer, are controlled  
 by the same parameter: the Reynolds number $\gamma$ (inverse of  $\alpha$-viscosity 
parameter), which in turn is related to the accretion rate 
 [see also \citet{tlm98}, hereafter TLM98].
It is worth noting that the identification of the break frequency as a diffusion effect (the inverse of  time
of the diffusion propagation in the bounded configuration) was later corroborated by both \citet{wo01}
and \citet{ga}. Particularly, Wood et al. demonstrated that the black hole candidate (BHC)  
XTE 1118+480  X-ray light curves with fast rise/exponential decay profile are a consequence 
of the diffusion matter propagation in the disk. On the other hand, Gilfanov \& Arefiev (2005) 
studied X-ray variability of persistent LMXBs in the $\sim 10^{-8}-10^{-1}$ Hz frequency range aiming 
to detect PDS features  associated with the  diffusion time scale of the accretion disk $t_{diff}$. 
As this is the longest intrinsic time scale of the disk, the power spectrum is
expected to be independent of the frequency below $\nu_b~(<1/t_{diff})$. They found that the break 
frequency correlates very well with the binary orbital frequency in a broad range of binary periods 
from $P_{orb}\sim 12$ min to  33.5 days, in accord with
theoretical expectations for the diffusion time scale of the disk.

\citet{zhang97} found while the low-energy X-ray (1.3-12 keV) and high-energy X-ray (20-200 keV) 
fluxes  strongly anticorrelate during the spectral transition, the bolometric luminosity in the soft states 
may only be  50\%-70\% greater than the hard state luminosity. On  the other hand, \citet{front01}  
found that the increase of the bolometric flux in the high/soft state with respect to that in the low/hard
 state is about s factor of $3$. 
In this Paper we further explore this issue of the bolometric luminosity using the data collected from 
the PCA and HEXTE detectors of {\it RXTE}. In fact, we confirm Zhang's et al. 
 finding that the bolometric luminosity slightly increases when Cyg X-1 undergoes transition from the 
 low/hard  to the soft states. We also comment on the issue of how the wind in Cyg X-1 affects the 
 bolometric luminosity.

\citet{pet78} and  \citet{kap98} argue that the X-ray source in Cyg X-1 is powered mainly by accretion from the strong stellar wind of the supergiant star. Cyg X-1
probably represents a situation intermediate between pure, spherical wind accretion and accretion 
by Roche lobe overflow. As \citet{gies} pointed out 
the density of the wind determines the size of X-ray ionization zone surrounding the black hole. This in turn controls the acceleration of the wind in the
direction of the black hole. During the low/hard  state, the strong wind is fast and accretion rate is relatively low, 
while during the soft state, the weaker,
ionized wind attains only a moderate velocity and the accretion rate increases.  It is evident that the Thomson optical depth of the wind 
increases in spectral transition because of decrease of the wind velocity (even if outflow mass rate is constant through the state transition).
We further investigate the effects of the wind in Cyg X-1 in terms of power and energy spectra, 
bolometric luminosities and the strength of K$_{\alpha}$ iron
line emission. 

The iron line observed in Cyg X-1 is the strongest among the
Galactic black holes. \citet{barr}  discovered the broad Fe K$_{\alpha}$ line, with equivalent
width $EW=120$ eV  and FWHM=1.2 keV in an {\it EXOSAT} spectrum of Cyg X-1. Because of these 
features are broad one should be concerned  
that the profiles are artifacts of inadequate continuum models or  instrumental effects. 
Miller et al. (2002) argue that an Fe K$_{\alpha}$ line is required
to obtain statistically acceptable fits to spectra observed from Cyg X-1 with a number of instruments, 
for a variety of continuum models and source luminosities
[see \citet{ebi}, and \citet{cui98} for {\it ASCA} results, and  see \citet{di01} and \citet{front01} 
for {\it BeppoSAX} results]. It is important to emphasize that most of the observations indicate that the 
relatively strong broad Fe K$_{\alpha}$ line emission which EW is within a range of 100-300 eV while
Miller et al. claim their best-fit model for Chandra spectrum includes a broad line ($E\sim 5.8$ keV, 
FWHM$\sim1.9$ keV, $EW=170\pm 70$ eV) component along with 
a narrow Gaussian emission line ($E\sim 6.4$ keV, FWHM$\sim 80$ eV, $EW=16\pm 3$ eV) component. 
It is clear that the inferred values of line energy E,  FWHM and  EW 
 are affected by the energy resolution of a given instrument and by the continuum model applied [see e.g. \citet{front01}]. Recently \citet{mh} revealed a strong and broad 
 Fe K$_{\alpha}$ line in another black hole source, GRS 1915+105 ($EW\sim 150-360$ eV). They suggest that there should be a link between EWs and  QPO frequencies in BH sources.  
Here we show that indeed there are  correlations between the strength of the iron line, QPO frequency and the spectral index.  
 
The observational signature of the mass accretion rate $\dot M$ is the QPO low frequency as it has
 been shown for BH sources by TF04.  The QPO frequency is related not only to  $\dot M$ but also 
 to the size of the Comptonizing region, $R$, i.e. $\nu_{QPO}\propto 1/R$. The behavior of $\nu_{QPO}$
  with respect to spectral index $\Gamma$ connects the characteristics of the Comptonization and spectral state with $\dot M$. This is graphically represented for BHs in the observations of \citet{vig}.
   We similarly employ this type of analysis to compare Cyg X-1
spectral states with other BHs to show their qualitative differences.

In \S 2 we present the details of the our spectral and timing  data analysis of archival RXTE data 
from the BH source Cyg X-1. In \S 3 we present and discuss the results of the data analysis and we
 compare them to that presented by TF04  for other BH sources.  In \S 3 we also offer an explanation 
 various correlations found in Cyg X-1.    Discussion and conclusions follow in \S 4.

\section{Observations and data analysis}

For our analysis we used data from Proportional Counter Array (PCA) and
High-Energy X-ray Timing Experiment (HEXTE) onboard {\it RXTE}. The data is available through 
the GSFC public archive \footnote{http://heasarc.gsfc.nasa.gov}.
Cyg X-1 is one of the sources  most extensively  observed by   {\it RXTE}. We searched 
the entire archive for public data.
The summary of the {\it RXTE} observation proposals and data used in the 
present analysis, and a reference to the corresponding proposal IDs, are given in Table 1. 
Each proposal consists of a set
observations that can be divided into intervals of continuous on-source 
exposure (usually about 3 ks) corresponding to one {\it RXTE} orbit. For 
each proposal we provide  its archival identification number (proposal ID), 
the dates between which the data were collected, the total on-source exposure,  the
number of continuous data intervals $N_{int}$,  the average number of operational 
PCUs in the proposal ${\bar{N}_{PCUon}}$.
The data spans $\sim 8$ years of data with almost
2.2 Ms of the total on-source exposure. 
We calculated an energy spectrum and an averaged PDS for each continuous
interval of data, which correspond to one 
orbital {\it RXTE} revolution. Data reduction and analysis was conducted with FTOOLS 5.3
software according to  recipes in the  ``{\it RXTE} Cook Book''.

\subsection{Timing Analysis}

Before being transmitted to ground-based station the PCA data is preprocessed
by six event analyzers, two  of which are always operating in Standard1 and Standard2 
modes.  
 To avoid telemetry overload for very bright sources such as Cyg X-1 
 the counts from several  energy
ranges are processed by separate on-board event analyzers and stored in separate
data files. During  most observations, Binned mode with several millisecond 
time resolution combining counts from 0 to 35 PCA energy channels is available.
Depending on RXTE epoch, this channel range corresponds to energy
range changing from 1.5-10 keV to 2.0-15 keV.
We were mostly interested in low frequency range ($\leq$ 100 Hz) and the 
temporal resolution of this mode is sufficient for our analysis.
Otherwise, we used sub-millisecond resolution Event or Single Bit Mode data.
The data was rebinned  
to a $2^{-11}$ second time resolution to obtain a Nyquist frequency of
1024 Hz. PDSs are normalized to give rms fractional variability per Hz.
 For the PDS modeling we used the 
broken power law \citep[see][for definition]{VS00}  
component to fit broad band frequency noise  
and Lorentzians to describe QPO profiles.

\subsection{Spectral Analysis}

Spectral data reduction and modeling was performed using the 
XSPEC astrophysical fitting package. First we performed the data
screening to calculate good time intervals for Fourier analysis.
We excluded the data collected for elevation angles less than 
$15^\circ$ and during South Atlantic Anomaly passage.
To avoid the electron contamination we also applied 
the condition for electron rate in the PCU 2 (which is operational during
all observations) to be less than 0.1 counts/sec. We extract energy spectra from Standard2 data files 
using counts from upper xenon layer of all operational detectors. 
 Then we applied a deadtime correction to account for detector
dwell time after each event detection. Current response
matrices for PCA and HEXTE give 10\%-20\% offset for relative cross-normalization.
Henceforth, when we fit PCA and HEXTE spectra
simultaneously, we  multiply the physical model applied to the 
data by a constant factor to account for this instrumental discrepancy.
 We fix the factor value at unity for PCA data set while allowing it to 
change for the HEXTE Cluster A and B spectra. In Table \ref{statestab} we
give the values of the best fit model parameters 
along with the offset values obtained for representative
PCA/HEXTE spectrum for each source state.

To describe
the continuum spectrum we use the Bulk Motion Comptonization ({\it BMC}) model 
which is a generic Comptonization model. This model can be used if the photon energy is less than the mean electron 
energy of the Compton cloud $E_{av}$.
 The choice of a particular 
theoretical model is provided by robust nature of the BMC model
for different spectral states and independence of a specific type of
Comptonization scenario involved. The BMC model spectrum  is a sum of
the  blackbody component (which is the disk radiation directly seen by the observer)
 and the fraction of the blackbody component Comptonized in the corona
 with the variable Comptonization fraction. The model has four parameters: $kT$ is a  color 
temperature of thermal photon spectrum,
 $\alpha$ is the energy spectral index ($\alpha=\Gamma-1$,  where $\Gamma$ is the photon index),
the parameter $A$  is related to  the weight of the Comptonized component, $A/(1+A)$,
and a normalization of the blackbody component. The BMC model is valid for the general 
case of Comptonization when both bulk and thermal motion are included.

For the thermal Comptonization $E_{av}$ is related
to electron temperature $E_{av} \sim kT_e$. When the bulk motion Comptonization is dominant
$E_{av}$ is related to the bulk kinetic energy of the electrons $E_{av}\lax m_e c^2$. 
For the thermal Comptonization for energies less than $E_{av}$ CompTT\footnote{CompTT model was introduced by Titarchuk (1994), Titarchuk \&
Lyubarskij (1995) and Hua \& Titarchuk (1995) to describe the spectrum of thermal Comptonization
in the whole energy range.} and
BMC models are identical. The thermal Comptonization and dynamical (bulk motion) Comptonization are
 presumably  responsible for the spectral formation
in the hard state and the soft state respectively. Therefore, one needs a generic 
 model such as  BMC that describes the spectral shape regardless of the specific type of Comptonization. 
Although the thermal Comptonization model can properly fit the spectral shape  for the observed spectra 
for all spectral states but it would give physically unreasonable values of
the best-fit parameters for the soft state spectra.  
Notably, Wilms et al. (2005) show optical depth of Compton Cloud inferred from CompTT
model significantly decreases towards high/soft state. It is very difficult (in the framework of any reasonable physical model) to
explain tendency when the  mass accretion rate increases during the hard-to-soft state transition. 

We remind a reader that TSh05 used two BMC components   
to fit the spectral data from  NS source  4U 1728-34.  
The radiation from the central object and inner parts of accretion disk is 
Comptonized by a surrounding cloud. As long as the spectral index  for each BMC component is a 
physical characteristic of the Comptonizing region, TSh05 used this parameter  for each of BMC 
components describing disk and NS Comptonized radiation. In fact,  the BMC XSPEC model spectrum 
(Titarchuk et al. 1997)  is a sum of the (disk or NS)  black-body component and Comptonized 
black-body component.
In the BH case one can conclude that only one the BMC component  is needed 
in order to describe  the Comptonization spectrum.
The soft photons generated by the solid surface are absent in this case.

\begin{figure}[ptbptbptb]
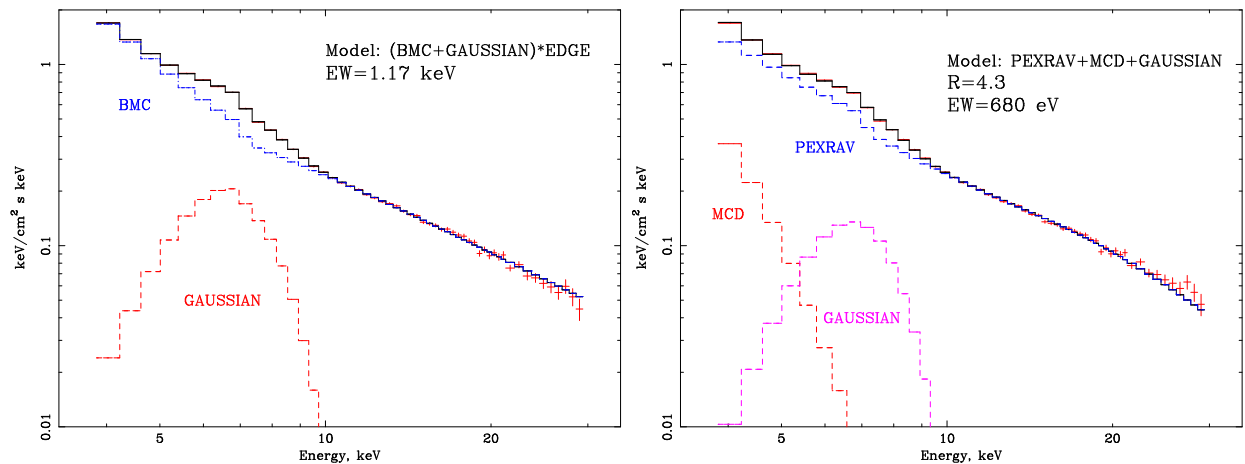

\includegraphics[scale=0.35,angle=-90]{f1a.eps}
\includegraphics[scale=0.35,angle=-90]{f1b.eps}
\caption{Comparative fits with BMC and PEXRAV models for the observation
during the high/soft state.}
\label{bmc_vs_pexrav}
\end{figure}

Another important issue in modeling the BHC spectra is a proper account  for
iron fluorescence line at $\sim6.4$ keV. As it was already mentioned, 
the iron line observed in Cyg X-1 is one of the strongest among BHC sources and is
confirmed by various instruments. The site of the iron line origin is 
not yet identified and the process of its formation is under debate. 
Currently it is fashionable to explain the iron line in the framework of  the reflection model by 
\citet{mz} (hereafter MZ95, see PEXRAV and PEXRIV models in XSPEC).  
The geometry required by the model implies 
that the source of  Comptonized component (exponentially cut-off power law) is located above 
a flat reflecting surface (accretion disk). The appearance of 
the iron line and so called a reflection hump at 15-20 keV is 
 then produced  as a  result of Compton reflection. Despite the fact that reflection model
provides a good spectral fits to the data, the best-fit parameters 
obtained for a soft state are hard to explain (we provide more details in the Discussion section).
In Figure \ref{bmc_vs_pexrav} we compare the fits given by BMC and PEXRAV models
for the observation taken when the source was in the soft state. The Reflection model fit 
requires an additional gaussian to account for the iron line. The line profile produced by
PEXRAV alone can not account for the total iron line strength even for high values of a reflection 
factor $R$. The large ``reflection'' component, in turn, leads to overestimated 
 spectral indices for the Comptonized component. 

\begin{figure}[ptbptbptb]
\includegraphics[scale=0.35,angle=-90]{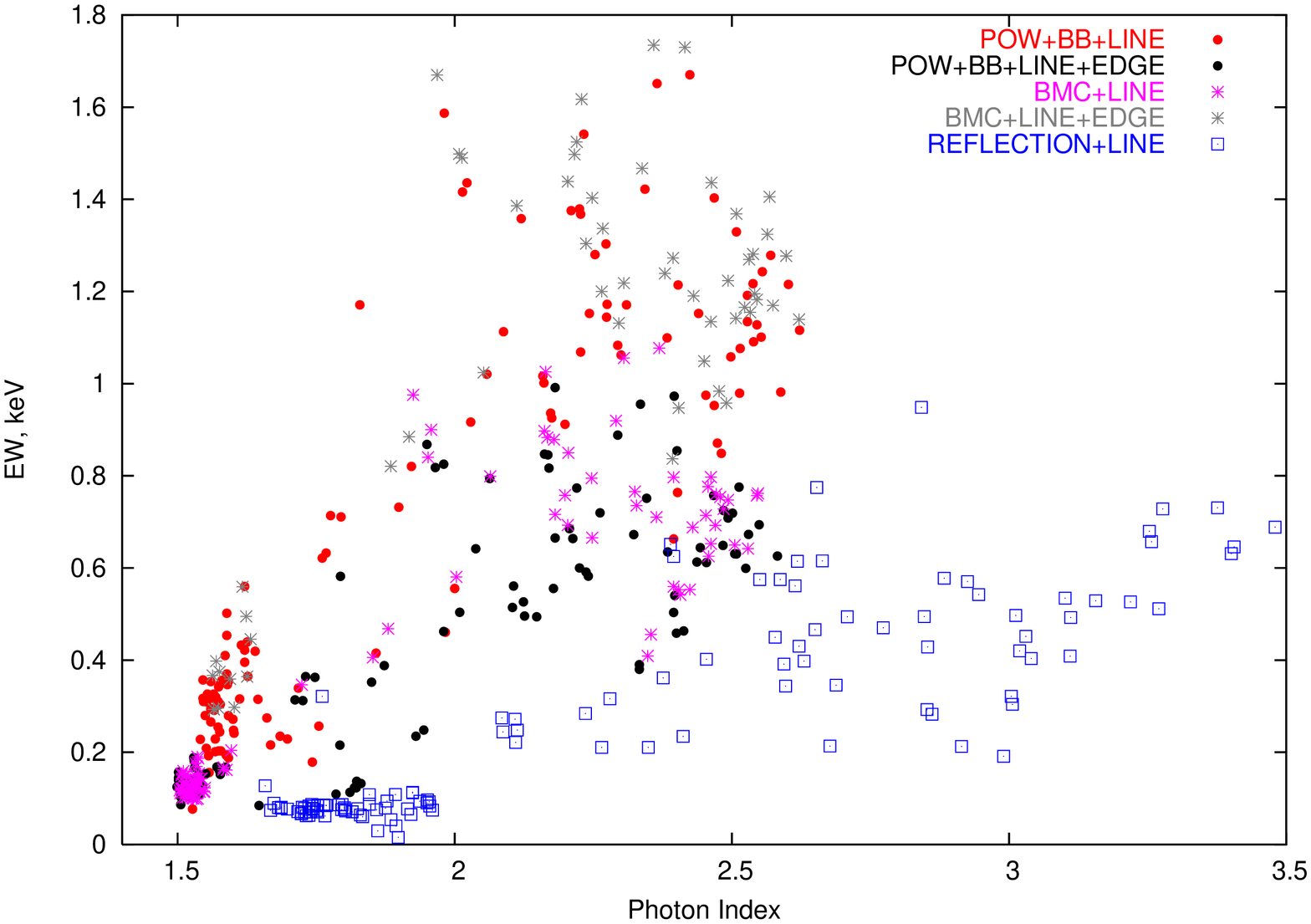}
\includegraphics[scale=0.35,angle=-90]{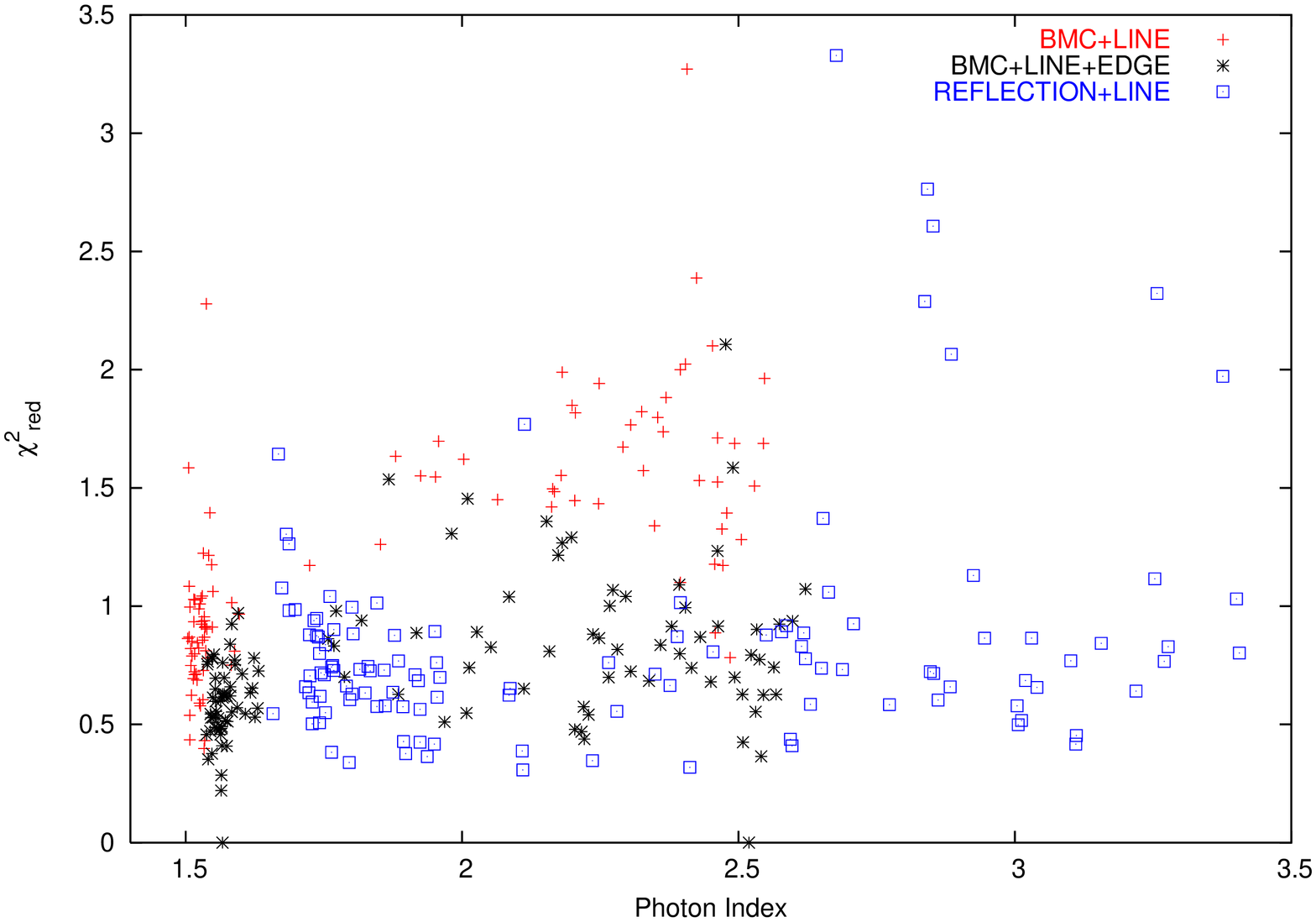}
\caption{Performance of different XSPEC models in a subset of Cyg X-1 RXTE data. Left hand panel for EW vs photon index
and right hand panel for $\chi^2$ vs photon index}
\label{models}
\end{figure}

\begin{figure}[ptbptbptb]
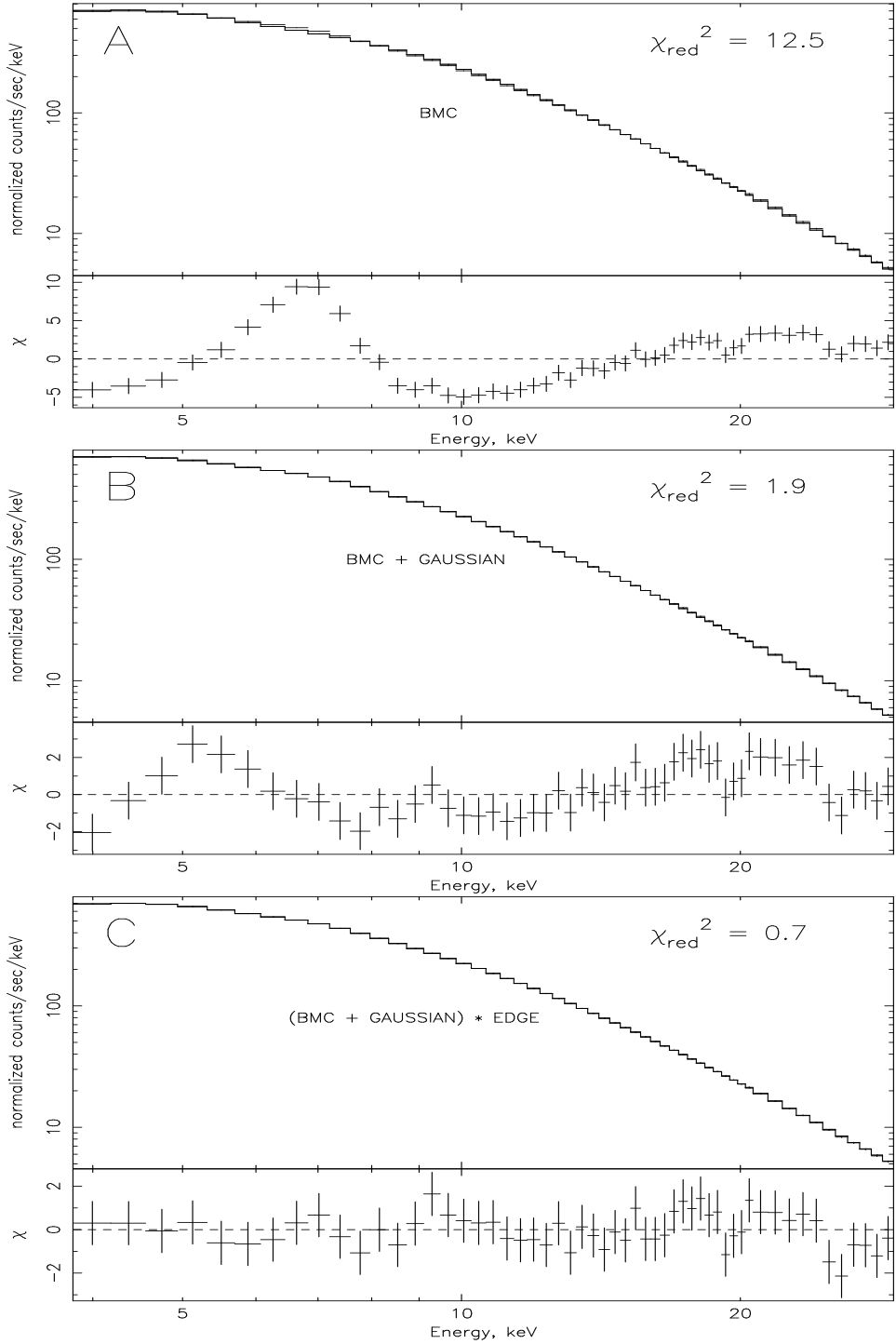

\includegraphics[width=2.5 in,height=5.in, angle=-90]{f3a.eps}
\includegraphics[width=2.5 in,height=5.in, angle=-90]{f3b.eps}
\includegraphics[width=2.5 in,height=5.in, angle=-90]{f3c.eps}
\caption{Consistency of the presence of  three components in X-ray spectra of Cyg X-1.  
Spectral fits by BMC model (upper panel), BMC+GAUSSIAN model (middle panel),  (BMC+GAUSSIAN)*EDGE model (lower panel).
}
\label{counts}
\end{figure}

An alternative mechanism for the line formation is proposed by \citet{lat04}, LaT04 hereafter.
In their scenario the iron line is formed in outflowing wind material. 
It is important to note that an absorption edge is {\it physically} required to 
compensate for absorption of photons above the K-threshold  energy [see \citet{kall04}] .
Our results favor this
model as it is consistent with equivalent width (EW) - spectral index correlations and
is capable of accounting for high EW values. To illustrate our arguments and to
investigate the mutual dependence of predicted spectral properties on the specific model
we fit the subset of Cyg X-1 observations which includes low/hard, soft states and 
transitions between them. The models and the results of fits are presented on Figure \ref{models}. 
The power law plus blackbody empirical model and BMC model provide qualitatively
similar results when the gaussian is used with the edge and without it. Considering the restricted
RXTE energy resolution at low energies the exact values of gaussian EWs have to be handled
with care. However, statistically the edge is highly significant.  
In Figure \ref{counts} we demonstrate the consistency of the our spectral model using  a
representative spectrum from Cyg X-1.
For all fits we fix  and relate the line energy at 6.4 keV to K threshold energy at 7.1 keV 
[see details of this relation in \citet{kall04}]. To account for interstellar absorption we use a fixed 
hydrogen column of $N_H=0.5 \times 10^{22}$ cm$^{-2}$.

Finally, according to the above arguments,  the XSPEC model for the spectral fitting reads as 
{\it WABS(GAUSSIAN+BMC)*EDGE}. 
We use $\alpha$, the  disk color temperatures, disk blackbody normalizations 
as free parameters of the spectral continuum model.  The spectral fits was obtained using 3.5-30.0 keV energy range.
We add 0.5\%  error to the data to account for systematic uncertainty in the  PCA 
calibration. The typical quality of fit is good with $\chi^2_{red}$ in the range of 0.5-1.5. 

\section{Data Analysis Results and The Inferred Model Parameters}
\subsection{The  evolution of the energy spectra in Cyg X-1}

\begin{figure}[ptbptbptb]
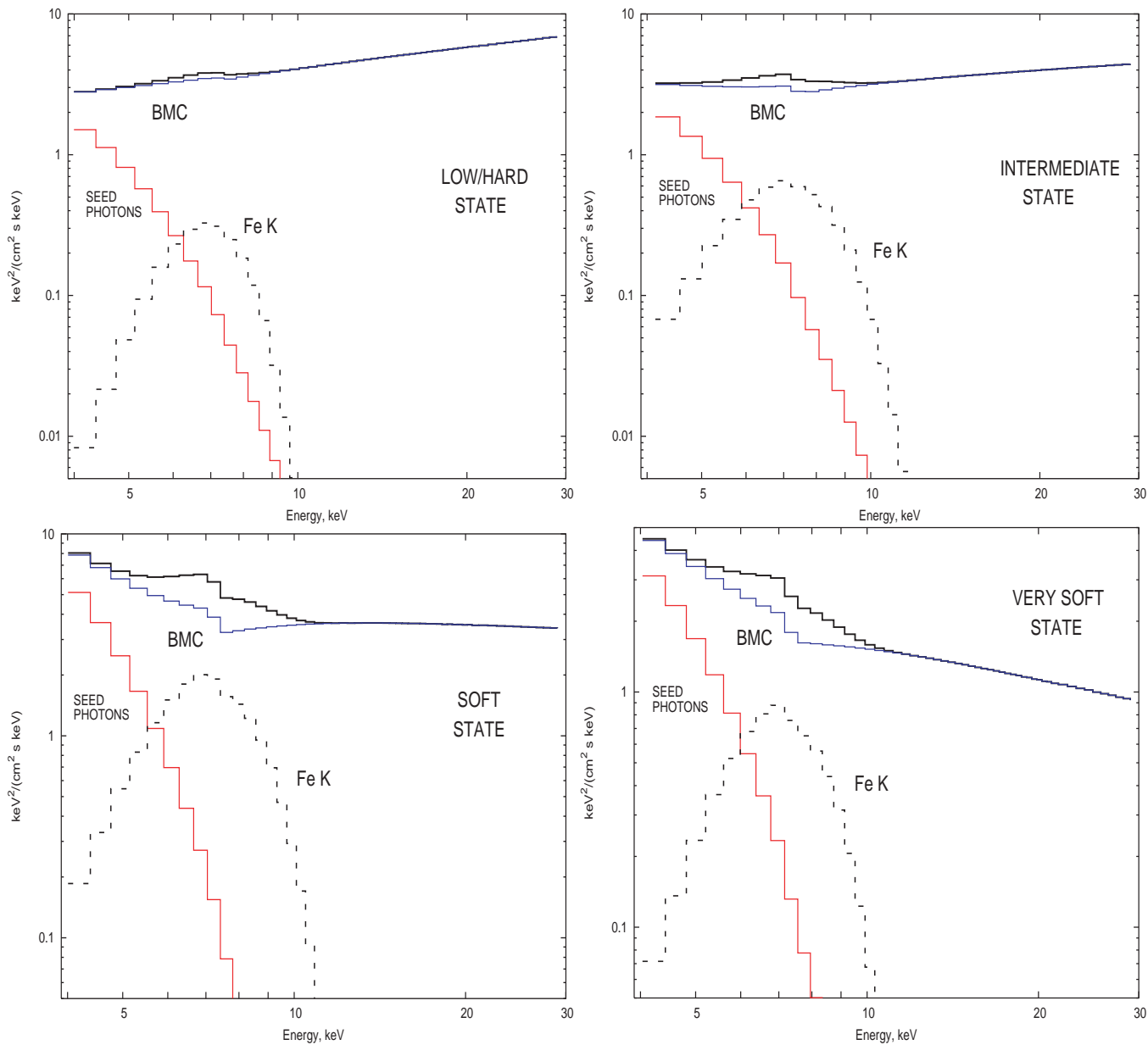

\includegraphics[width=3.5 in,height=3.2in,angle=0]{f4a.eps}
\includegraphics[width=3.5 in,height=3.2in,angle=0]{f4b.eps}
\includegraphics[width=3.5 in,height=3.2in,angle=0]{f4c.eps}
\includegraphics[width=3.5 in,height=3.2in,angle=0]{f4d.eps}
\caption{
Spectral evolution of the source from  low/hard state (panel a), through intermediate state (panel b) and  soft state (panel c) to
very soft state (panel d). Photon index of the upscattering
Green function $\Gamma$ changes from 1.5 to 2.6 respectively.    
}
\label{spevol}
\end{figure}

\begin{figure}[ptbptbptb]
\includegraphics[width=5 in,height=6in,angle=-90]{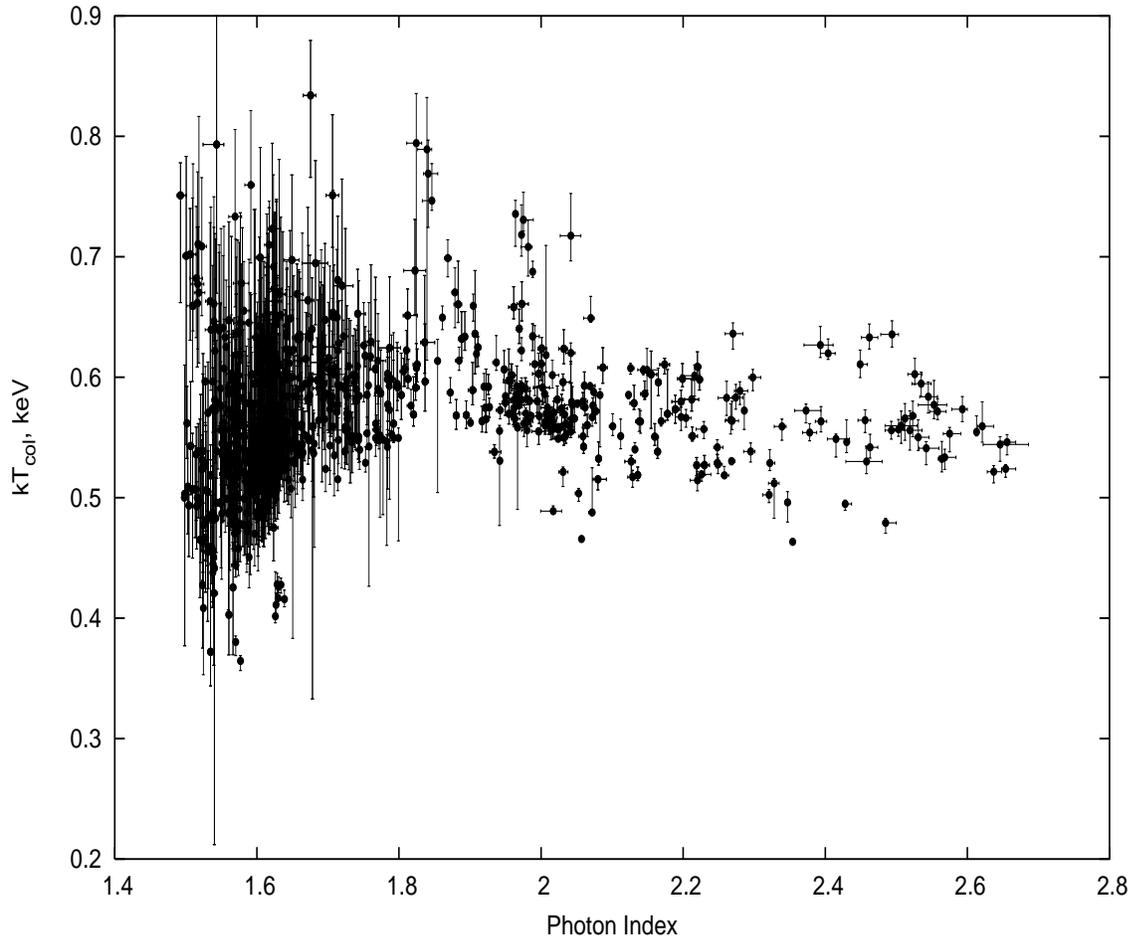}
\caption{
Color temperature of thermal (disk) component vs. photon index. Low/hard state corresponds to the left hand side of the picture; 
soft states correspons to the right hand side. }
\label{kT}
\end{figure}

The evolution of spectral properties of the source during the transition from the low/hard to 
the soft states is shown on Figure \ref{spevol} and in Table \ref{statestab}.  The temperature profile 
of  the   thermal (blackbody) component is plotted versus the index $\Gamma$ in Figure 
\ref{kT}.  This  temperature is presumably related to the disk. 
It changes only slightly  in the narrow range of 0.5 -0.6 keV and is nearly independent of the spectral state.
 
 When the source is in the low/hard state the emerging radiation spectrum is presumably
formed as a result of Comptonization of soft photons generated in the disk. 
We present a $\nu F_{\nu}$- diagram for the low/hard state in upper left hand panel of 
Figure \ref{spevol} ($F_{\nu}$ is the energy  flux).  As a source progresses to higher luminosity 
states the spectrum becomes softer, the power-law part of the spectrum becomes steeper and the 
contribution of the blackbody (thermal) component increases. The  strength of the iron 
line also increases (see  the right upper panel in Figure \ref{spevol}). One can explain this evolution 
of the spectrum by an increasingly efficient deposition of the gravitational energy in the disk which becomes stronger towards the soft states \citep[see e.g.][]{ct95,esin}.

The luminosity reaches the highest level in the soft state when the spectral index $\Gamma$  
is about 2. The power-law plateau  is clearly seen in $\nu F_{\nu}$- diagram presented in the lower 
left hand  panel of Figure \ref{spevol}. Then, as the spectral index increases we observe a slight decrease
in flux. We identify this phase as a very soft state (the so called `` thermal dominated state"), in addition to the canonical low/hard, intermediate and 
soft states.  The energy spectrum is dominated by
thermal component (more than 70\% of total flux) and the maximum observed spectral index is
$\sim2.7$. It is important to emphasize that, typically, 
the high/soft state in the BH sources is observed when an extended power-law component has index 
$\Gamma\sim2.7$ [see e.g. \citet{grove}, \citet{bor}, hereafter BRT99, TF04]. 
 This relatively low index for Cyg X-1 soft state can be explained by a higher temperature of the 
converging flow than that for the
 high/soft state of other BH sources \citep[see][hereafter LT99, and also TF04]{lt01}. 
 The main reason for this may be the fact that in the soft state of Cyg X-1, the  energy  release 
in the disk and in the Compton cloud are comparable while in the typical LMXB  
BH sources, like GRS 1915+105, XTE J1550-564, GRO 1655-40, in the high/soft state
  the energy release in the disk is much greater. When spectral 
index progresses to values higher than $\sim2$ the luminosity decreases (see the lower 
right hand panel of Figure \ref{spevol}). 

It is worth noting that we use a terminology  for the spectral states in Cyg X-1 based on our physical 
scenario of the spectral evolution there (see more of the details in the discussion section). In 
our classification the difference between soft and very  soft states corresponds to the difference 
in the spectral indices. In  soft state when  the saturation of the spectral index $\Gamma$ 
vs QPO frequency $\nu_L$ occurs,  $\Gamma$ is about 2 (see \S 3.4). Whereas in the very soft state 
the spectra become softer and  $\Gamma$ increases to the values of 2.7. No QPO frequencies are 
observed in this state.  
 
 In the scheme of McClintock and Remillard (2004) (and every other "high" or "very high state" BHC 
sources) this classification can look different.  One should be careful in applying the definition of 
spectral states (particularly soft states) for some particular source. 

\subsection{The  evolution of the  power spectra in Cyg X-1} 

The next important question is how this spectral evolution is related to the timing characteristics of the
source. In our  study of the power density spectrum (PDS),   we reveal that the PDS features,  
break frequency $\nu_b$, and Lorentzian low-frequency $\nu_L$ and Q-value of the QPO frequency evolve and 
increase while the 
source progresses toward the soft state. But  the QPO frequencies  are completely washed out in the 
very soft state. \citet{tcw}, hereafter TCW02, predicted that when the source is embedded in the optically 
thick  medium the QPO features must be absent in the PDS of the source because of photon scattering. 

When the radiation from the central source passes through the surrounding cloud of optical depth 
$\tau_0$ and of radius $R$ the direct (unscattered)  and scatterred  fractions of the radiation
are  $\exp(-\tau_0)$ and [$1-\exp(-\tau_0)$] respectively. 
Consequently the rms amplitude  of the direct component decreases exponentially  with $\tau_0$.    
TCW02 show that the rms amplitude of the scattered component $B(\omega)$ for a given rotational frequency $\omega=2\pi\nu$   decreases as
\begin{equation}
B(\omega)\propto \exp(-2\chi),
\label{bomega}
\end{equation}
where $\chi(\omega, \tau_0)=\{[\pi\tau_0/2(\tau_0+2/3)]^4 +(3R\tau_0\omega/4c)^2\}^{1/2}$.

As seen from equation (\ref{bomega}), the QPO amplitude of the radiation scatterred in the  wind 
(cloud)   decreases very rapidly with  $\tau_0$ and with $\omega$ ($\nu$), namely 

\begin{equation}
\chi \sim 4.7\left(\frac{\tau_0}{1}\right)\left(\frac{R}{10^{11}~{\rm cm}}\right)\left(\frac{\omega}{2\pi\times 0.3~{\rm Hz}}\right)\gg 1.
\label{argum}
\end{equation} 

A radius R of $10^{11}$ cm was chosen as a typical radius of the wind inferred from the observations [\citet{gies}] and theoretically inferred 
using luminosity and spectrum as observable characteristics  of the source [\citet{lat04}]. 

Thus we can conclude that the variability of the scatterred part of radiation is completely washed out  in the wind of radius of order $10^{11}$ cm even for
$\tau_0\gax0.5$. The variability of the direct (unscattered) component is preserved but its rms amplitude  decreases as $\exp(-\tau_0)$  with $\tau_0$.
 
The power spectrum analysis of Cyg X-1 data confirms this expectation (see Pottschmidt et 2003; Axelsson, 
Borgonovo \& Larsson 2005 and our results in this paper). 
The power spectrum in the very soft state 
is featureless (see Fig. \ref{power}). The emission of the central source is presumably obscured by the optically thick wind and consequently 
all photons emanating from the central source are scattered. The direct component of the central source radiation that carries information about the
variability is suppressed by scattering. \citet{gies} argue that there is a particular state of Cyg X-1 when the wind velocity is very low and thus one can expect high accumulation
 material in the wind and high optical depth of the wind.  The wind downscattering of the photons 
emanating from the inner Compton cloud leads to the softening of spectrum \citep{tsh05} 
and consequently to a decrease in X-ray luminosity of the source. The softening of the spectrum 
can also be a result of effective cooling the Compton cloud by the disk soft photons.
If the power law component of the soft state is formed in the converging flow, then the 
high indices are a result of  the low flow temperature (LT99). 
The index  increases and  saturates to the critical value about 2.8 with the mass accretion rate  for the low temperatures of the flow (see LT99 and Titarchuk \& Zannias 1998, hereafter
TZ98).  

To calculate the power spectrum of the central source emission affected by scattering one should present the quantitative model of the resulting pulse affected by 
 scattering $z(t)$ which can be written as follows:
\begin{equation}
z(t)=\int_0^t x(t^{\prime}) g(t-t^{\prime})dt^{\prime},
\label{conv}
\end{equation}
where $x(t)$ is the input pulse of the central source oscillations and  $g(t)$ is the response pulse of the scattering medium.
The power spectrum of $z(t)$ is a product of the power spectrum of $x(t)$ and $g(t)$, i.e.
\begin{equation}
||F(\omega)||^2=||F_{cs}(\omega)||^2||F_{scat}(\omega)||^2,
\label{pwsp}
\end{equation}
where   
\begin{equation}
F(\omega)=\int_0^{\infty}e^{-i\omega t}z(t)dt,
\label{ftz}
\end{equation}

\begin{equation}
F_{cs}(\omega)=\int_0^{\infty}e^{-i\omega t}x(t)dt,
\label{ftx}
\end{equation}

\begin{equation}
F_{scat}(\omega)=\int_0^{\infty}e^{-i\omega t}g(t)dt
\label{ftg}
\end{equation}
are  Fourier transforms of $z(t), ~x(t),~G(t)$ respectively.  
The time response of the innermost part of the source $x(t)$ 
can be considered in terms of diffusion propagation of perturbation in
the disk transition layer (TL). Any local perturbation in TL (bounded medium) would propagate diffusively outward over time scale (see TL99) 
\begin{equation}
t_{dif}\sim \frac{l_{fp}}{v}\left(\frac{L}{l_{fp}}\right)^2=\tau_{pert}\frac{L}{v},
\label{cstime}
\end{equation}
where $L=R_{out}-R_{in}$ is the characteristic thickness of the TL,  $l_{fp}\sim \eta/(\rho v)=(\sigma_{pert} n)^{-1}$ is the mean free perturbation path related to the turbulent MHD viscosity $\eta$, the matter density $\rho$, the number density $n$ and the perturbation interaction 
 cross-section $\sigma_{pert}$ in the TL and $\tau_{pert}=L/l_{fp}=\sigma_{pert} n L$. 
 Even though the specific mechanism providing this viscosity needs to be understood, this time scale $t_{dif}$ apparently ``controls'' the diffusion supply of the matter into the
 innermost region of the accretion disk (TL).

The response function $x(t)$  is a solution the time-dependent problem for perturbation diffusion that mathematically is formulated as an initial value  problem for the
diffusion equation in the bounded medium (see more details of the diffusion theory in ST80, ST85, T94).  This solution is a linear combination  
\begin{equation}
x(t)= \sum_{k=1}^{\infty}A_ke^{-\lambda_k^2t/3t_{fp}},
\label{xtdiff}
\end{equation}
where $t_{fp}=l_{fp}/v$ is a free path crossing time,  $t_{k,diff}=(\lambda_k^2/3t_{fp})^{-1}$ is kth eigen diffusion time,  $A_k=c_kX_k(R_{out})$, $c_k$ is 
expansion coefficient of the source perturbation function $f(r)$ 
\begin{equation}
c_k=\frac{\int_{R_{in}}^{R_{out}}\varphi(r)X_k(r)f(r)dr}{{\int_{R_{in}}^{R_{out}}\varphi(r)X_k^2(r)dr}}
\label{expcoef}
\end{equation}
that is related to eigenvalues $\lambda_k$ and eigen functions $X_k(r)$ of the appropriate space diffusion operator and $\varphi (r)$ is the operator weight function.
 
If $\tau_{pert}\lax 1$   then the response function as a solution of the diffusion problem can be presented by a single exponent, namely 
\begin{equation}
x(t)\approx A_1e^{-\lambda_1^2t/3t_{fp}}=A_1e^{-t/t_{cs}},
\end{equation}
because  in this case $\lambda_1^2/3\sim 1/\tau_{pert}^2\lax 1$ and $\lambda_1^2/3\ll\lambda_k^2/3$, for $k=2,3,...$, 
where $t_{cs}=t_{1,diff}=(\lambda_1^2/3t_{fp})^{-1}$ 
(ST80, ST85, T94).  

The scattering response function $g(t)$ is a solution of the  diffusion problem for the photon scattering in the extended envelope (wind)  that surrounds the central source.
The formula for the photon diffusion time $t_{scat}$ is very similar to formula (\ref{cstime}), i.e.
\begin{equation}
t_{scat}\sim  \tau_{0}\frac{L_w}{c},
\label{scattime}
\end{equation} 
where $L_w=R_{w,out}-R_{w,in}$ is the envelope thickness and the Thomson optical depth of the envelope $\tau_0=\sigma_{\rm T} n_e L$.
When $\tau_0\lax1$  (similarly that for the perturbation diffusion) the scattering diffusion function $g(t)$ can be presented by the exponent
\begin{equation}
g(t)\approx A_{1,w}e^{-t/t_{scat}}.
\label{gtdiff0}
\end{equation}
It is worth noting that $t_{scat}\gg t_{cs}$, because $L_w\gg L$ by definition of $L_w$ and $L$.

For these particular response functions (see Eqs. \ref{xtdiff}, \ref{gtdiff0}) the power spectra are
\begin{equation}
||F_{cs}||^2\propto (\omega^2 + t_{cs}^{-2})^{-1},
\label{cspower0}
\end{equation}
\begin{equation}
||F_{scat}||^2\propto (\omega^2 + t_{scat}^{-2})^{-1},
\label{scatpower0}
\end{equation}
and
\begin{equation}
||F||^2=||F_{cs}||^2||F_{scat}||^2 \propto (\omega^2 + t_{cs}^{-2})^{-1}(\omega^2 + t_{scat}^{-2})^{-1}.
\label{power0}
\end{equation}
Because $t_{scat}\gg t_{cs}$   we have the following asymptotics for the power spectrum $||F||^2(\nu)$ as a function of $\nu~(=\omega/2\pi$)
\begin{eqnarray}
\label{pwasym1}
||F||^2(\nu)\propto (t_{cs}t_{scat})^2\nu^0~~~~~~~~~~~~~~~~~~~~~~{\rm for}~~ \nu\ll 1/(2\pi t_{scat})\\
\label{pwasym2}
||F||^2(\nu)\propto (t_{scat}/2\pi)^2 \nu^{-2}~~~~~~{\rm for}~~ 1/(2\pi t_{scat})\ll \nu\ll 1/(2\pi t_{cs})\\ 
\label{pwasym3}
||F||^2(\nu)\propto(2\pi)^{-4} \nu^{-4}~~~~~~~~~~~~~~~~~~~~~ {\rm for}~~  \nu\gg 1/(2\pi t_{cs}).
\end{eqnarray}
In Figure \ref{power} one can clear see the change of the power-law index of the $\nu\times$power diagram from $(0)$  to $(-2)$  about 0.08, 0.5, 3 Hz  in the low/hard, intermediate and
 the  soft  states respectively.
Thus one conclude that the perturbation diffusion times in the TL (Compton cloud) $t_{dif}\sim 1/(2\pi\nu_{b}$ Hz) are about $2,~0.3,~0.05$ s in these  states.

The change of the power-law index of the diagram from (1) to (-1) [that
corresponds to 0 -(-2) changes of that in the power spectrum (Eqs. \ref{pwasym1}, 
\ref{pwasym2})] takes place  in the low-frequency parts of the diagrams for the low/hard, intermediate and soft states (black, blue, red respectively). 
The power at high frequencies decays very fast and the corresponding power-law index of PDS is about 4 (see Eq. \ref{pwasym3}). 

However  in the $\nu\times$power diagram for the very soft state  the index changes from 0 to (-1) at low-frequency (about 0.2 Hz) 
but this change is not described by aforementioned formulas 
(\ref{power0}, see also \ref {pwasym1}, \ref {pwasym2}).  We argue this particular power-law index transition occurs in the source when the optical depth 
of the wind $\tau_0\gg1$.

In order to demonstrate this effect one must calculate the power spectrum of the scattering response 
function $g(t)$ for  $\tau_0\gg1$. The derivation of these formulas are out of scope of the paper and 
we shall present these results elsewhere. 

It is worth commenting that we have already shown here, the resulting
power spectrum  is a product of two power spectra related to the disk PDS and another one 
to Compton cloud (scattering) PDS (see Eq. \ref{pwsp}). In the general case, each of these spectra 
is a power spectrum of the series of eigen exponential shots [see for example, formula \ref{xtdiff} for $x(t)$]. 
For frequencies much less than the characteristic scattering frequency, $\nu\ll\nu_{scat}=1/(2\pi t_{scatt})$
the resulting spectrum is represented by a power-law red noise component. In our simple treatment of this problem
considered here and applicable to the hard and intermediate states only (see Eq. \ref{power0}) the low-frequency power-law index of PDS is zero. 
In fact, the index and power-law cuttoff frequency increases with mass accretion rate when the source undergoes transition to a soft state.
We shall provide all  details of this correlation in a future publication.

\subsection{Correlation of K$_{\alpha}$ line strength with the photon index}

Another observational appearance of the wind is the strong broad feature of K$_{\alpha}$ line in
the spectrum that is present in all spectral states of Cyg X-1 (see Fig. \ref{spevol}). In Figure \ref{line_index} (upper panel) we demonstrate how equivalent
width EW of the  K$_{\alpha}$ line increases with the photon
index $\Gamma$ from about 150 eV in the low/hard state to about 1.3 keV in the high/soft and very soft states.
 One can see signs of saturation  of the EW 
at about 1.3 keV   for indices above 2.1 when no QPO is present in the power spectrum.  
Wilms et al. (2005) have also found that one needs a strong  K$_{\alpha}$ line for fitting data with a 
Comptonization model (CompTT). The strength of the line increases almost linearly with a accretion disk 
flux until it satures at the values of $\sim 1$ keV. 

The X-ray photons presumably originated in the innermost part of source, in an area less than 40 
Schwarzschild radii, illuminate the wind. In this case the wind gas is heated by Compton scattering 
and photoionizations  from the central object. It is cooled by radiation, ionization,
and adiabatic expansion losses (LaT04). 
The photons above the K-edge energy are absorbed and  ionize iron atoms that leads to the formation of
the strong K$_{\alpha}$ line.  LaT04 calculated ionization, temperature structure and the equivalent widths
 of Fe K$_{\alpha}$ line formed in the wind. For the wide set
of parameters of the wind (velocity, the Thomson optical depth $\tau_{\rm T}$) and the incident 
Comptonization spectrum (the index and the Compton cloud electron temperature) they established 
that EW of the line should be about  1 keV and less for the line to  be observed. 
LaT04 also predicted that for this case the inner radius of the wind should be situated at 
$(10^{3}-10^{4})/\tau_{0}$ Schwarzschild radii away from the central object. 

In the framework of the wind model we can inferred  the optical depth of the outer shield (wind) 
as a function of the index using the EW of K$_{\alpha}$ line [see \citet{basko}, CT95]:
\begin{equation}
EW=\omega_K\int_{E_{th}}^{\infty}(E/E_{th})^{-\alpha}\{1-\exp[-(Y+Y_0)\tau_{0}(7.8~{\rm keV}/E)^{3}]\}dE.
\label{EW_tau}
\end{equation}
where $\omega_K$ is the fluorescence yield, $E_{th}$ is the K shell ionization threshold energy,   
$Y$ and $Y_0$ are the abundances of elements (in units of the cosmic abundances) with a charge 
$Z<26$ and the iron abundance, respectively.
This integral can easy be calculated using the following formula
\begin{equation}
EW=7.8~{\rm keV}\times\omega_K\sum_{m=1}^{\infty}(-1)^{m-1}\frac{[(Y+Y_0)\tau_0]^m}{(3m+\alpha-1)m!}\left(\frac{7.8~{\rm keV}}
{E_{th}}\right)^{3m-1}.
\label{EW_tau_form}
\end{equation}
We apply
the value of 
$E_{th} \sim 7.1$  keV 
[see \citet{kall04}], and $\omega_K=0.34$ [see \citet{bam}].
In Figure \ref{line_index} (lower panel) we present the inferred dependence of $\tau_0(Y+Y_0)$ on the photon index.

To obtain  a description of the X-ray photon spectrum of Cyg X-1 many  authors 
(see e.g. Gilfanov, Churazov \& Revnivtsev 1999, hereafter GCR99; P03)  used an empirical model in which each source spectrum is a sum of power law 
spectrum with photon index $\Gamma$ and a multi-temperature disk blackbody (Makishima et al. 1986).
 To this continuum, a reflection spectrum after MZ95 was 
added. GCR99 emphasized that this empirical model is obviously oversimplified and therefore the best-fit parameter values do not
necessarily represent  physically meaningful quantities. Particular problems arise with the values of the reflection factor 
$R=\Omega/2\pi$ and the equivalent width of the iron line.
 The best-fit values of $R=\Omega/2\pi$ for the soft state exceed the unity considerably $(R=1-5)$, which
is physically meaningless in the geometry of the reflection. In fact, Lapidus, Sunyaev \& Titarchuk (1985), 
showed that the maximum reflection factor for geometrically thin infinite disk illuminated by 
isotropic radiation from the central object is 0.25.

The values of the EW  of K$_{\alpha}$ inferred by GCR99 vary from 80 to 300 eV. These EWs are related to the
 line component included in the spectrum in addition 
to the refelection component.  Because the reflection model includes its own iron line component 
(MZ95) thus the actual strength of the line  is much higher in Cyg X-1 than that 
 presented in GCR99.  One can determine that the total iron line  strength strongly  
increases with  photon index because of the power of the so called ``reflection component'' R, which 
also strongly increases with the index (see Figs 5, 8 in CGR99). In this sense our inferred strong lines  
in the soft states of Cyg X-1 agree with those in GCR99. 
       
\subsection{The index-QPO correlation} 

In Figure \ref{ind_qpo} we present the correlation of photon index $\Gamma$   
versus low QPO frequency $\nu_L$ (orange points in the upper panel)  and the correlation of  $\Gamma$   
versus break frequency $\nu_b$ (the lower panel) observed in Cyg X-1.

 We compare the index-QPO correlation 
with those observed in GRS 1915+105, XTE J1550-564. 
One  notices similar properties  for all of these sources: i. when QPO is detected 
the photon index does not go above 3, ii. the index saturates. at low and high values of the QPO frequencies.
  In Cyg X-1 the saturation level of the index for high values of low QPO frequency  is remarkably 
lower than for the other two sources. 

TF04 argues that the index saturation level is determined by the temperature of the converging flow 
where the soft (disk) photons are upscattered by electrons to the
energy of falling electrons (LT99). In principle, one can evaluate the mass of the central BH using the 
index-QPO relation
 because QPO frequencies are inversely proportional mass (TF04).
The simple slide of the index-QPO correlation for XTE J1550-564 (pink line) over the frequency axis gives 
us  the index-QPO correlation for   GRS 1915+105 (blue line). 
The shifting factor is
10/12 which gives the relative BH mass in XTE J1550-564 with respect to that in GRS 1915+105. 
But one caveat should be taken into account: this sliding method works if the index-QPO relations are 
self-similar with respect to each other  as occurs  for GRS 1915+105 and  XTE J1550-564.   

\subsection{Flux-index and Comptonization fraction-index relation}

In Figure \ref{flux-index} we show how the 1-30 keV flux (black crosses) varies during the spectral transition from the low/hard state (L/H) to very soft (VS) state.
We use joint PCA/HEXTE spectral fits for bolometric corrections for high energies.
We extracted HEXTE spectra from Cluster A and Cluster B using the same screening criteria that we obtained for PCA spectra.
To fit PCA/HEXTE data we apply the model obtained for PCA multiplied  by high energy cutoff ({\it HIGHECUT}) component to account for high energy turnover in the hard tail of the  X-ray spectrum.   
The  flux  is then calculated by integrating the best-fit model spectrum over the energy interval
 from 1 keV to 300 keV. Absolute normalization of HEXTE data for both clusters
  were allowed to change free with respect to PCA normalization. The resulting fits
  show the HEXTE normalization consistently less than the normalization for PCA  by 15-20\%
as expected.  Resulting values of bolometric flux are shown  
on Figure \ref{flux-index} (red circles). 

We confirmed Zhang's et al. (1997) claims that the bolometric flux remains almost unchanged 
(within 50\%, almost) during L/H to VS transition.
This phenomenon can be explained as a combined effect of the mass accretion rate increase and the 
Comptonization enhancement decrease when the system undergoes
the spectral transition. In the low/hard state the relative small energy release (low mass accretion rate) in the disk is compensated by high Comptonization
efficiency in the corona while in the very soft (thermal dominated) state the situation is opposite.
Sunyaev \& Titarchuk (1980),  hereafter ST80, and  Sunyaev \& Titarchuk (1985),  hereafter ST85, derive  the asymptotic form of the Comptonization enhancement
factor ${\cal E}_{\rm Comp}$ for both regimes of the energy spectral index ($\alpha<1$ and  $\alpha\geq1$). 
Chakrabarti \& Titarchuk (1995), hereafter CT95, provides the general formula for ${\cal E}_{\rm Comp}$ (CT95, formula 14) which combines these two asymptotic.   
In order to infer  the bolometric flux dependence $F_{bol}$  on the  photon index using the observable thermal (disk) flux $F_{th}$ one should multiply
 $F_{th}$ by $[1+(A/(A+1)){\cal E}_{\rm Comp}]$ and apply formula (14) in CT95 for ${\cal E}_{\rm Comp}$, namely
 \begin{equation}
 F_{bol}= F_{th}\left[1+\frac{A}{1+A}{\cal E}_{\rm Comp}(\alpha, x_0)\right],
 \label{flux_bol}
 \end{equation}
where $x_0=2.7 kT_{col}/kT_e$,  $kT$ is a color temperature of disk radiation (see Fig \ref{kT}) and $kT_e$ is electron temperature of Compton cloud.
The enhancement factor ${\cal E}_{\rm Comp}$ depends on $kT_e$ only for $\alpha<1$ ($\Gamma<2$) (see formula 14 in CT95). 

In order to infer $kT_e$ we apply use the same   PCA/HEXTE data. 
To calculate the electron temperature $kT_e$ we  use the {\it HIGHECUT} parameter, $E_{fold}$   and well-known relation between the electron temperature and 
the high energy cutoff of the Comptonization spectrum, namely $kT_e\approx E_{fold}/2$ (see e.g. ST80 and T94). 
We present the results of calulations of ${\cal E}_{\rm Comp}(\alpha, x_0)$ as  a function of $\Gamma(=\alpha+1)$ for different values of $kT_{col}$ (see Fig. \ref{ecomp}).

In Figure \ref{flux-index} we  demonstrate a dependence of $F_{bol}$ on the photon index 
$\Gamma(=\alpha+1)$ (blue triangles). The theoretically predicted and observed
flux values are in good agreement along.    
Thus {\it the inferred luminosty-index relation  supports the idea that  
the variation of luminosity with index is due to the combined effect of the disk mass accretion rate and 
Comptonization of the disk photons in the corona}. 

The soft photon radiation is completely  Comptonized in the low/hard state while the relative contribution of the 
Comptonized radiation decreases toward the very soft
(thermal dominated) state. In Figure \ref{comp_ratio} we present the observed correlation between the ratio of the 
Comptonized flux to the bolometric flux and the index.
The Comptonized spectrum is a convolution of the soft photon spectrum with the upscattering Comptonization Green's 
function. This component of the observable spectrum and
consequently the related flux can be obtained using the BMC model. The inferred Comptonized fraction of the 
spectrum helps us to reveal the relative size of the corona 
(Compton cloud) with respect to the disk emission region. One can see that in the soft  states the coronal region 
becomes more compact. This effect is also confirmed
by the observed index-QPO correlations (see Fig. \ref{ind_qpo}). The QPO frequencies increases with the index  as a result of the mass accretion rate increase. 
In fact, the coronal  region
is pushed close to BH when mass accretion rate goes up (see TLM98 and TF04). In the  soft states (when the Compton cloud is relatively cold)
the converging flow site ($\lax 3-4 R_{\rm S}$)  is only the place  where the soft (disk) photons get scattered
 due to the dynamical Comptonization. 

\subsection{Break frequency- QPO low frequency relation and BH mass determination}

In Figure \ref{br_lowfr} we present the observed correlation between the break frequency $\nu_b$ and the
low frequency $\nu_L$. We fit this correlation by the broken power law of the form
\begin{equation}
\nu_b=a(\nu_L/\nu_{\ast})^{\beta_1}~~~~{\rm for}~~~~\nu_L<\nu_{\ast}~~~{\rm and}~~~~
\nu_b=a(\nu_L/\nu_{\ast})^{\beta_2}~~~~{\rm for}~~~~\nu_L\geq\nu_{\ast}.
\label{broken}
\end{equation} 
The best-fit parameters: the normalization $a\sim 0.32$, the ``low''-frequency index 
$\beta_1\sim 1 $ and the ``high''
frequency index $\beta_2\sim1.65$ and the ``boundary'' index $\nu_{\ast}=2.2 $ Hz. Titarchuk, Osherovich \& Kuznetsov
(1999) found that the ``high'' frequency index $\beta_2\sim 1.6$ is a canonical index of break-low frequency
correlation for a quite a few BH and NS sources. In fact, Titarchuk \& Osherovich (1999) identified using
dimensional analysis the corresponding radial oscillation   and diffusion frequencies in the transition layer
(TL) with the low-Lorentzian $\nu_{L}$ and break frequencies $\nu_b$ for 4U 1728-34. 
They predicted values for $\nu_b$ related to the diffusion in the transition layer, that are consistent
 with the observed $\nu_b$.
TO99 argue that $\nu_{L}$ is the inverse of the oscillation time of the TL radial mode, 
\begin{equation}
\nu_L = C_{bd}\frac{v}{L}\sim t_{QPO}^{-1}
\label{lowfr}
\end{equation} 
and  $\nu_{b}$ is the inverse of the diffusion time of the perturbation propagation in the TL
\begin{equation}
\nu_b =  \frac{v}{f\tau_{pert} L}\sim t_{dif}^{-1} ,
\label{break}
\end{equation}
where $L$ is the TL radial size (see definition  of $\tau_{pert}$ in the text after Eq. \ref{cstime}). $C_{bd}$ is the coefficient which depends on the specific outer and inner 
boundary conditions imposed in the TL (see, Titarchuk, Bradshaw \& Wood  2001, 
for details of the boundary problem of the diffusion and oscillations in TL).
Diffusion time $t_{dif}$ is related to the diffusion length $L_{dif}=l_{fp}N_{int}$ where $l_{fp}$ is a free
propagation path of the perturbation in the TL, $N_{int}=f(L/l_{fp})^2=f\tau_{pert}^2$ is a number of matter interactions, related to  
 the effective viscosity in the TL. Factor $f$ is related to the source distribution of the
perturbation in the TL (see Sunyaev \& Titarchuk 1985 for details).
Thus the relation between $\nu_b$ and $\nu_L$ reads as follows
\begin{equation}
\nu_b = \frac{1}{f\tau_{pert}} \nu_{L}.
\label{break-low1}
\end{equation}    
Because the perturbation dimensionless depth $\tau=L/l_{fp}$  is related to $L$  then 
\begin{equation}
\nu_b = \frac{1}{f\tau_{pert}} \nu_{L}=\frac{C_{bd}l_{fp}v}{fL^2}.
\label{break-low2}
\end{equation} 
Thus  one can find taking into account Eqs.(\ref{lowfr}, \ref{break-low2}) that $\nu_b$ should not be  a linear function of $\nu_L$ if 
$f\tau_{pert}\gg1$.
TO99 found that $\nu_b\propto \nu_{L}^{1.6}$ when $\tau_{pert}\gg 1$. The number of interactions in the TL hydrodynamical  flow $N_{int}$  and 
the perturbation  depth $\tau_{pert}=L/l_{fp}$ depends strongly  on the mass accretion rate.  For relatively high rates one should expect that 
$N_{int}\gg1$ and $f\tau\gg1$ while these values are about one for relatively small mass accretion rates that occur in the low/hard state.
In the latter case the dependence of $\nu_b$ on $\nu_L$ is almost linear. 

This diffusion effect is  confirmed by the correlation of $\nu_b$
and $\nu_L$  observed in Cyg X-1 (see Fig \ref{br_lowfr}). For higher values of frequencies $\nu_b$ ans $\nu_L$ (which correspond to higher values of mass accretion rate)
$\nu_b\propto \nu_{L}^{1.6}$ while for lower values $\nu_b\propto \nu_{L}$. 

The low frequency $\nu_L$ is inversely proportional to the TL size which its turn is proportional to the mass of the central object $m=M/M_{\odot}$ (BH or NS). Thus  one can conclude
that $\nu_L$ and $\nu_b$  are inversely proportional to m  when mass accretion rate in a source 
is relatively small. This inverse proportionality of $\nu_b$ vs $m$ can  be used for the mass determination of the objects that mass differs 
by order  of magnitude from Galactic BHs for example, of the supermassive or intermediate BHs  provided the break frequency is detected there.
Recently, Fiorito \& Titarchuk (2004)  applied  rescaling  of QPO frequency $\nu_L$  to evaluate a BH central mass in ultraluminous source
M82 X-1 while  McHardy et al. (2005) and Dewangan, Titarchuk \& Griffiths (2006) applied  rescaling of $\nu_b$  for the BH mass determination in AGN and ULX respectively.

\section{Conclusions and Discussion}

We have presented a detailed spectral and timing analysis of
X-ray data for Cyg X-1 collected with the {\it  RXTE}.
We find observational evidence for the correlation of
spectral index with low-frequency features: $\nu_b$ and  
$\nu_{L}$. The photon index $\Gamma$ steadily increases from 1.5
in the low/hard state to values exceeding 2.1 in the soft state. 
The low frequency $\nu_L$ is detected throughout the low/hard and intermediate states, while it disappears when the source
undergoes transition to the very soft (thermal dominated) state.
Like in other BH sources, there is  an indication of saturation
 of the index in the soft state for Cyg X-1 (see Fig.\ref{ind_qpo}). This saturation 
 effect, which  is presumably due to photon trapping in the converging flow, 
can be  considered to be a BH signature.  We want to stress that this saturation  is a model
 independent phenomenon found in the present analysis of X-ray {\it RXTE} data.
On the other hand, {\it the saturation of the index with mass accretion rate increase (which is strongly 
related to the QPO frequency increase)
has to apply to any BH, because the photons are unavoidably trapped in the central accretion flow. 
It is a necessary condition of BH presense in the accreting systems}. 

\begin{figure}[ptbptbptb]
\includegraphics[width=6 in,height=6in,angle=-90]{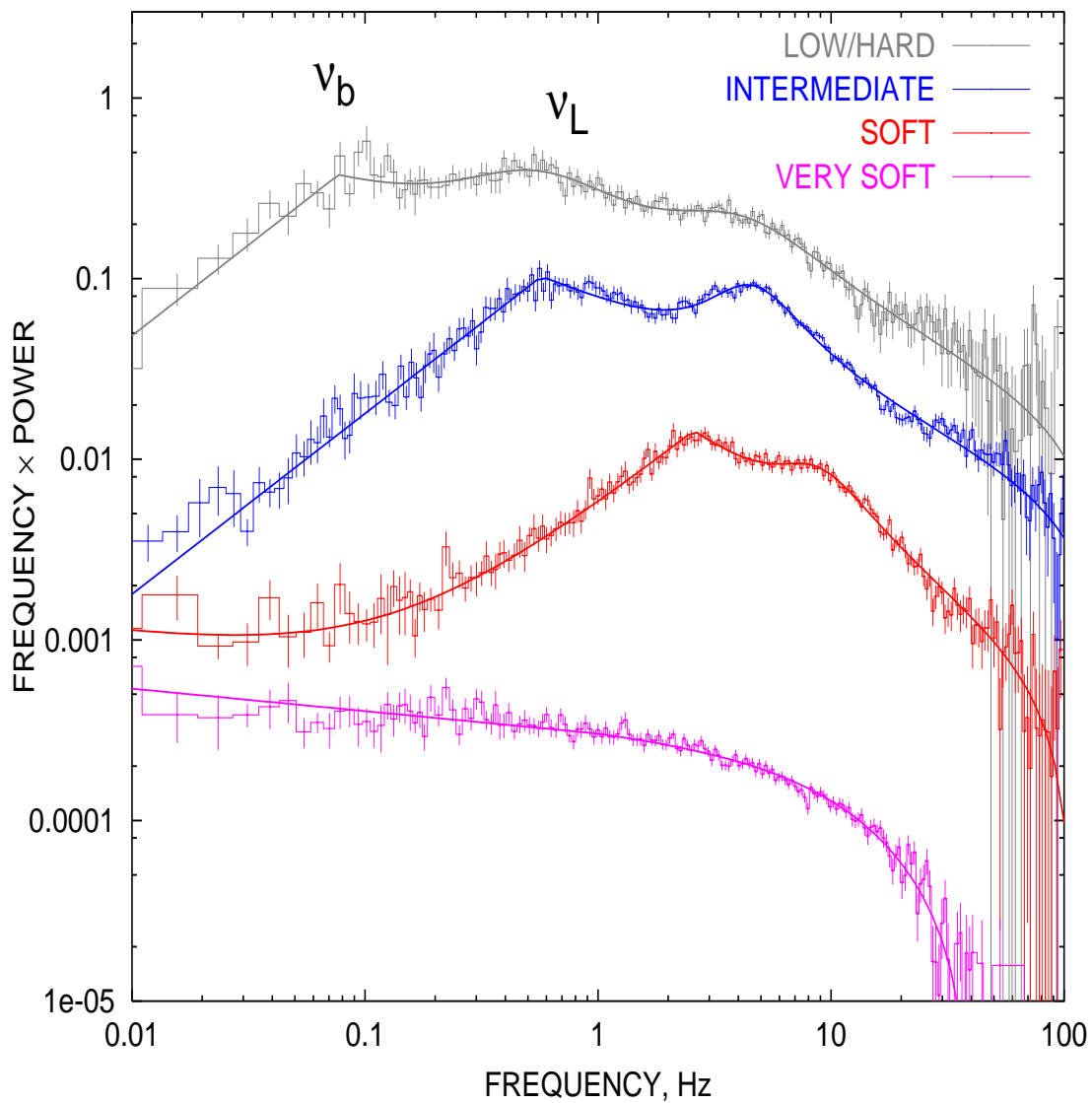}
\caption{
Power spectrum (PDS) evolution of the source from  low/hard state (black) through intermediate state (blue) and
soft state (red) to very soft state (purple). The break frequency $\nu_b$ and the low QPO frequency are clearly seen in 
low/hard, intermediate state and soft state. In the very soft state the power spectrum is featureless, no QPO and break are present. 
Any break and QPO features are washed out. }
\label{power}
\end{figure}

Thus the index saturation with QPO frequency seen in the source (rather than the presense of the tail in the soft state) 
is a signature of the horizon. In fact, one can see high energy tails with indices at about the BH
 saturation value 2.8 in the high/soft state observations of weakly magnetized accreting NS binaries, 
for example GX 17+2 (Farinelli et al. 2005)  and 4U 1728+34 (TSh05). Farinelli et al. (2005) presented two spectra of GX 17+2 observed in 1997 by BeppoSAX. Using these spectra
alone one cannot establish the evolution of the indices with the mass accretion rate. 
TSh05 analyzing $\sim 1.5$ Ms of {\it RXTE} archival data for 4U 1728-34 reveal the spectral
evolution of the Comptonized blackbody spectra and  QPO frequencies when the source transitions 
 between hard to soft states. Contrary to the BH sources, 
the indices of 4U 1728-34 spectra  do not saturate as QPO frequency increases. 
They increase from $\sim 2$ (in the hard state) to $\sim6$ (in the soft state) with no signature 
of saturation versus QPO frequency (or mass accretion rate). The NS soft state spectrum 
consists of two blackbody components that are only slightly Comptonized (inferred photon indices 
of the Comptonization Green's function are $\Gamma\gax6$). Thus one can claim (as expected from theory) 
that in NS sources thermal equilibrium is established for high mass accretion rate (soft) state.
In BHs the equilibrium is never established because of the presence of the event horizon. 
The emergent BH spectrum, even in the soft state, has a power-law component which
 index saturates with mass accretion rate. It is worth noting that there is a particular state in 
BH source when it can show signs of the thermal  equilibrium: the emergent spectrum consists of one 
or two thermal components. But no QPO is observed then.
In this (very soft, thermal dominated) state the source is presumably covered by a powerful wind that 
thermalizes the radiation of the central source and prevents 
to see any QPO generated in the source (see Fig. \ref{power}).     
   
One can argue that the bulk motion Comptonization (BMC) is ruled out as a main radiative process 
in the soft spectral states of black-hole binaries because of the
inefficiency of producing photons with energies $\gax 100$ keV and the lower relative normalization 
of the BMC component (see a recent paper by 
Niedzwiecki \& Zdziarski 2005 on Monte Carlo simulation of the bulk motion Comptonization, 
hereafter NZ05). The production of high energy photons  in Monte Carlo 
simulation with steep power-laws ($\Gamma>2$) is a  technically  difficult 
problem because of poor statistics. It requires a long  simulations and special methods 
for the   treatment of the  poor statistics at high energies. 
LT99 implemented this technique and found in their  simulations that the spectra extend up
to 200-300 keV. In fact, NZ05 do not show  histograms of their simulated spectra
but rather show the best-fit curves  describing  results of their simulations. It is 
impossible to quantify uncertainties and the quality of counting 
statistics as a function of energy in the NZ05 simulations. 

 The absolute normalization of the Comptonized component is always  determined  by the 
 seed photon normalization (illumination pattern).
LT99 and then Turolla, Zane \& Titarchuk (2001) and NZ05 confirm that the illumination 
pattern does not affect the  shape of the BMC spectrum. 
Thus, the relative normalization of the Comptonized (BMC) with respect to that of blackbody 
 is a model dependent
parameter. LT99 show that the BMC normalization can be  very high and comparable 
with the disk BB component when the innermost part the accretion flow 
(BMC area of 1-3 $R_{\rm S}$) is illuminated by the soft photons coming from the 
geometrical thick disk situated very close. The normalization is quite low
when the seed photons  illuminated the converging flow region come from the 
geometrically thin disk (see NZ05).           

In contrast to the claim by NZ05, which is based on their spectral modeling,  
we find that  Cyg X-1  observations show a very strong  Bulk Motion Comptonization 
signature  in the soft state as a photon index saturation with QPO frequency (mass accretion rate).  
This  is  a well-known signature of the photon trapping in the converging flow
discovered by TZ98 and then confirmed in Monte Carlo simulations by LT99.

\begin{figure}[ptbptbptb]
\includegraphics[width=3.8in,height=6.5in,angle=-90]{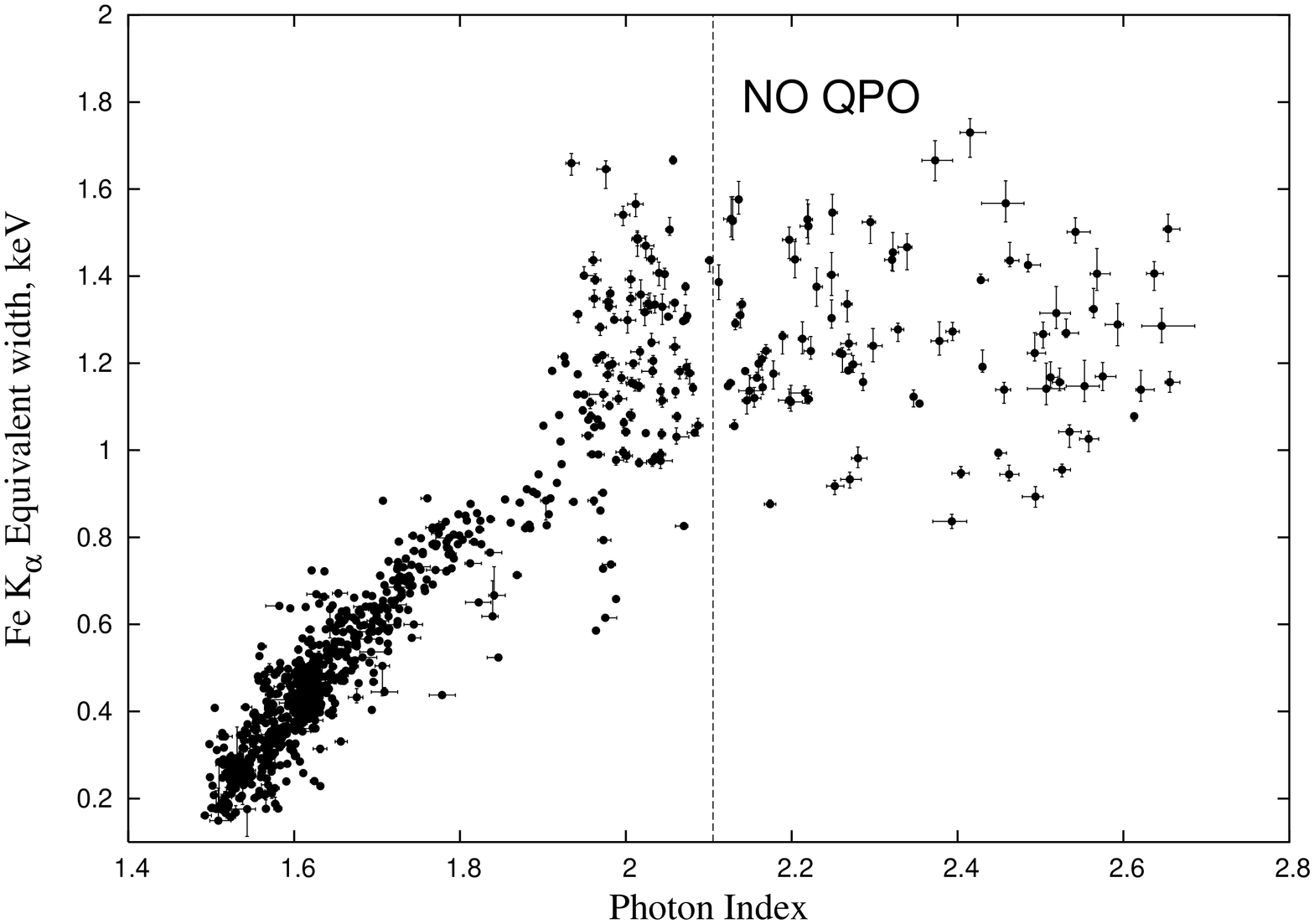}
\includegraphics[width=3.8in,height=6.5in,angle=-90]{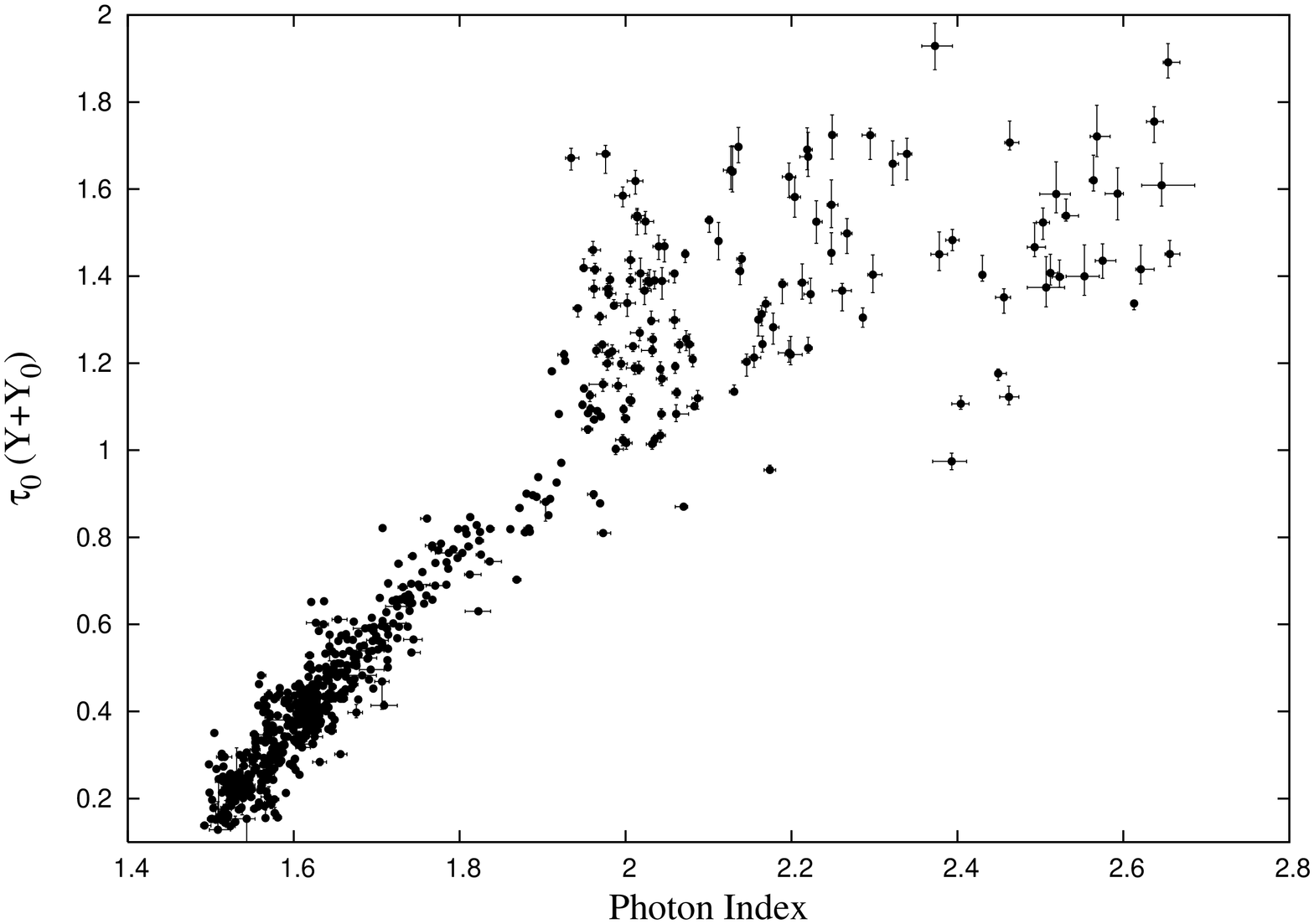}
\caption{
Upper panel: observed  Fe K$_{\alpha}$ line equivalent width (EW) versus photon index  $\Gamma$. 
EW increases toward the soft state. It is   $\sim 150$ eV in the low/hard state
and $\sim 1.3$ keV in soft and very soft states. 
Lower panel: inferred  product of $\tau_0(Y+Y_0)$ versus  $\Gamma$.
 }
\label{line_index}
\end{figure}

\begin{figure}[ptbptbptb]
\includegraphics[width=3.75in,height=6.2in,angle=-90]{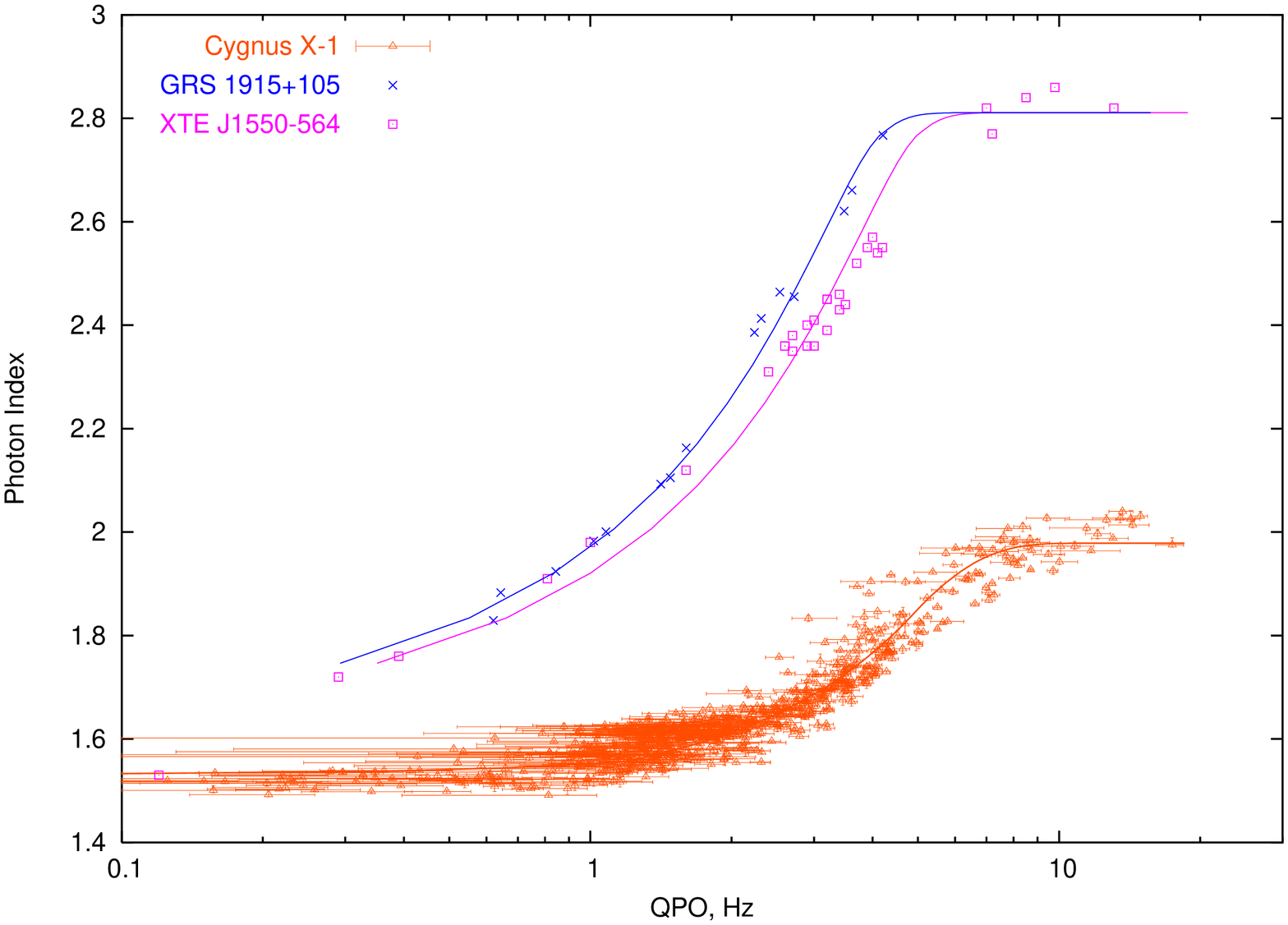}
\includegraphics[width=3.75in,height=6.2in,angle=-90]{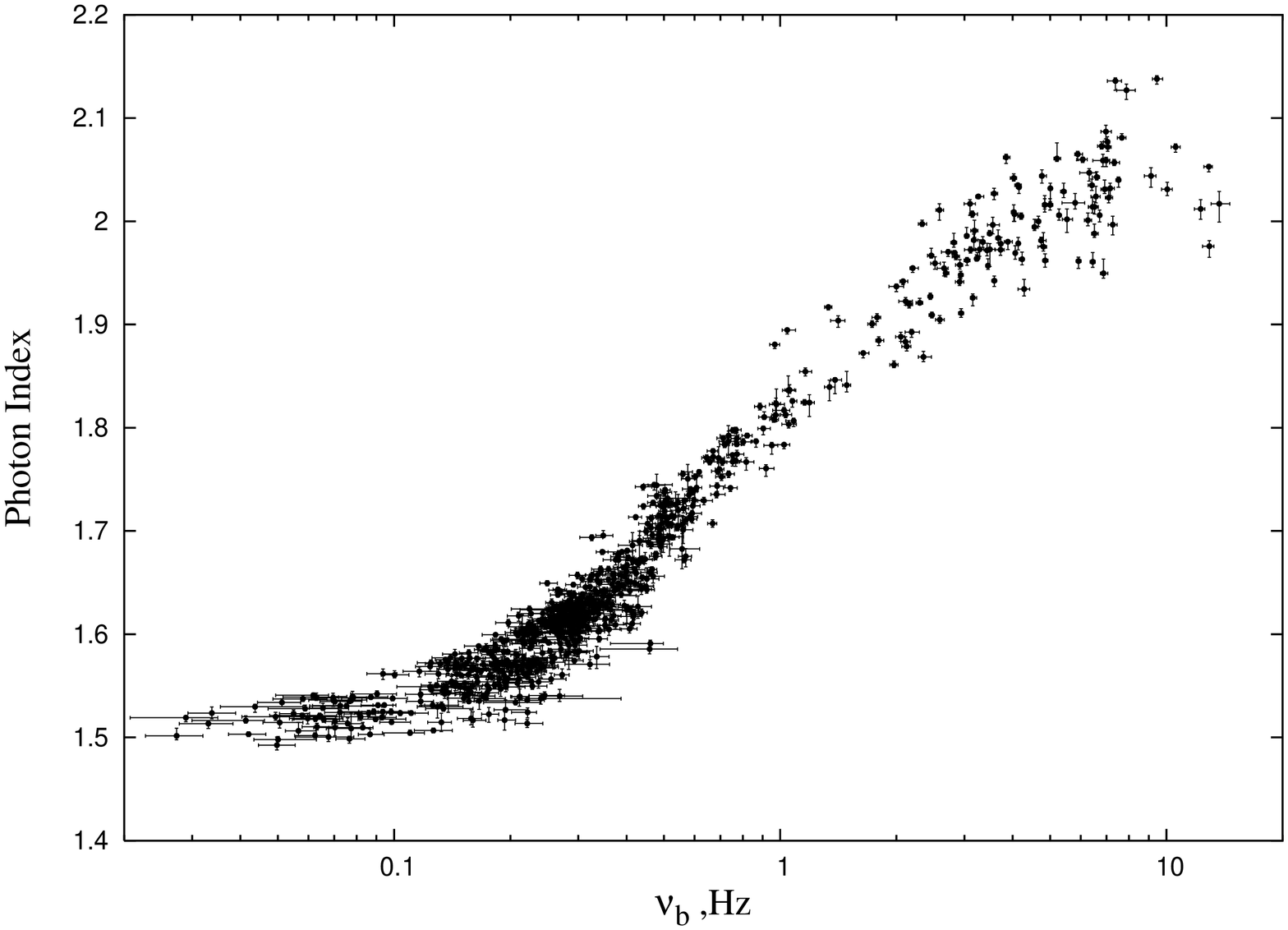}
\caption{
Upper panel: The observed  correlations
 between photon index $\Gamma$ and low frequency $\nu_L$ (orange points) which are compared with that in two other BHCs sources, XTE
 J1550-564 and GRS 1915+105. The saturation value of the index varies from source to source but it does not exceed the
 theoretically predicted value $2.8$ for the converging flow of  non-relativistic temperature (see TZ98).
 Presumably, the saturation value of the index depends on the plasma temperature of the converging flow (LT99).
Lower panel: The observed  correlation
 between photon index $\Gamma$ and break frequency $\nu_B$ for Cyg X-1}
\label{ind_qpo}
\end{figure}

\begin{figure}[ptbptbptb]
\includegraphics[width=4.in,height=6.1in,angle=-90]{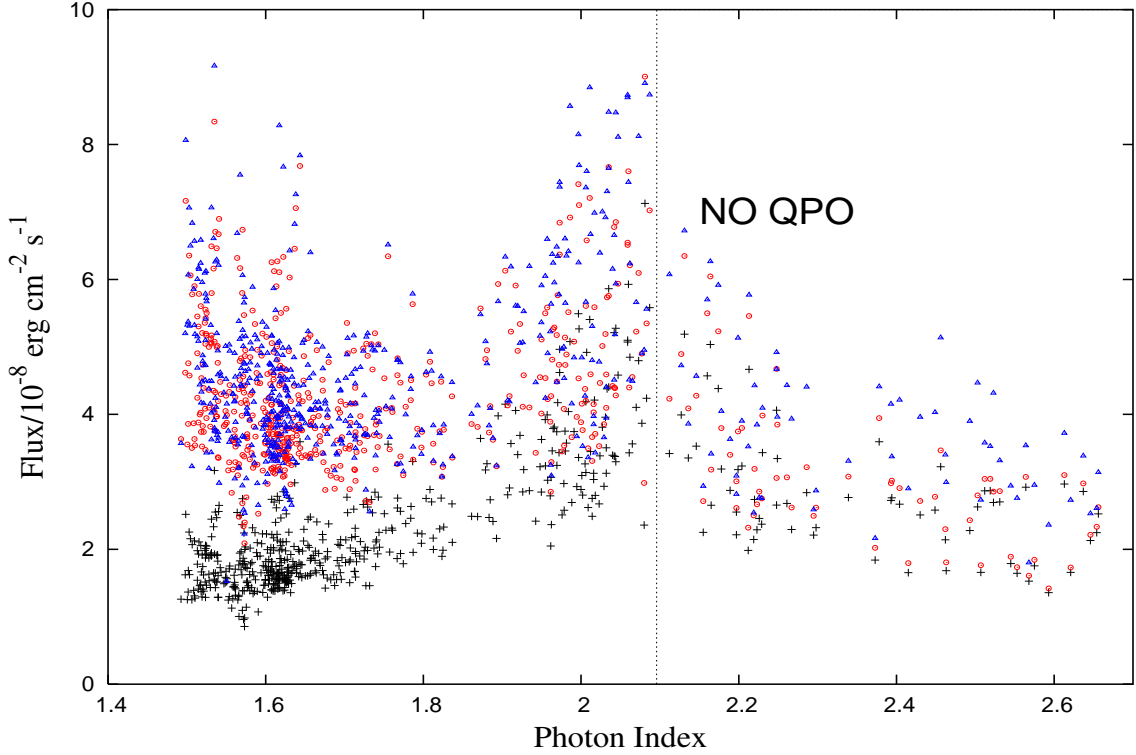}
\caption{
Flux in the energy band from 1 keV to 30 keV (black crosses) as a function of the index $\Gamma$.
Bolometric corrected flux (1-300 keV) is shown in red circles. Blue triangles is a total flux inferred from the Comptonization
enhancement factor ${\cal E}_{\rm Comp}$. The total fluxes given by bolometric correction and Comptonization model
show excellent consistency.
The 1-30 keV flux gradually increases toward the soft state with ($\Gamma\sim 2.1$), 
while the total flux stay almost constant, only slightly increasing during this soft state. 
In the very soft state ($\Gamma> 2.1$, no QPO detected) the flux sharply drops with the index 
presumably due to downscattering  of X-ray emission of the central source in the
strong wind.}
\label{flux-index}
\end{figure}

\begin{figure}[ptbptbptb]
\includegraphics[scale=0.65,angle=-90]{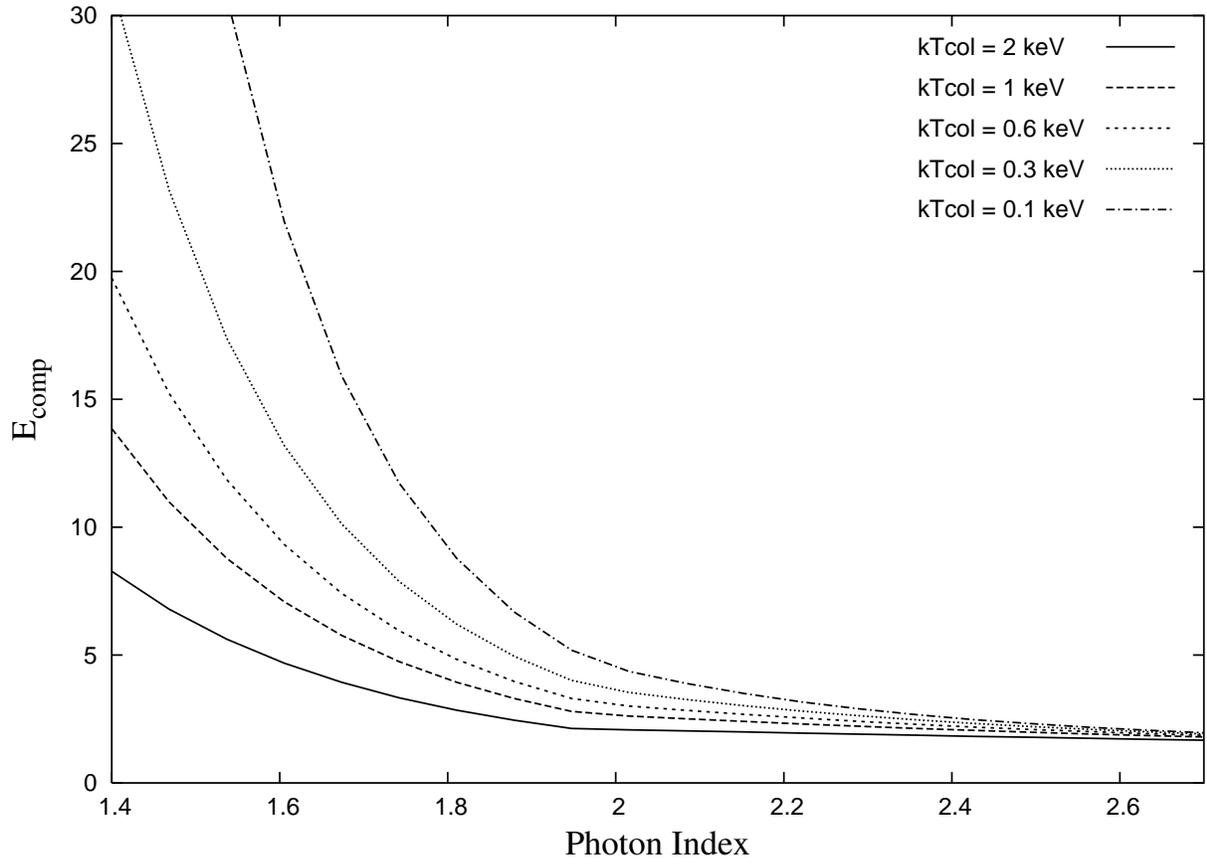}
\caption{ Comptonization enhancement factor ${\cal E}_{\rm Comp}$ vs photon index $\Gamma$.}
\label{ecomp}
\end{figure}

We also demonstrate that 
the Fe K$_{\alpha}$ line equivalent width  correlates   with spectral index and correspondingly with 
 QPO frequencies when they are present in the data (see Figs. \ref{line_index}, \ref{ind_qpo}).
This leads us to conclude that the compactness of the X-ray  emission area 
(taking the QPO frequency value as a compactness indicator) is higher for
softer spectra (related to higher mass accretion rate). On the other hand the 
Fe K$_{\alpha}$ emission-line strength (EW) is about one keV when the power spectrum is featureless.
 It happens in the very soft (thermal dominated)  state.  
Thus the observations may be suggesting that the photospheric radius of the Fe K$_{\alpha}$ emission is
orders of magnitude larger than that for the X-ray continuum. We propose that  Fe K$_{\alpha}$ 
line emission originates in the wind where the photons 
emanated from the central part of the source are downscattered by electrons and absorbed 
by partly ionized iron atoms. 

\begin{figure}[ptbptbptb]
\includegraphics[width=4.in,height=6.1in,angle=-90]{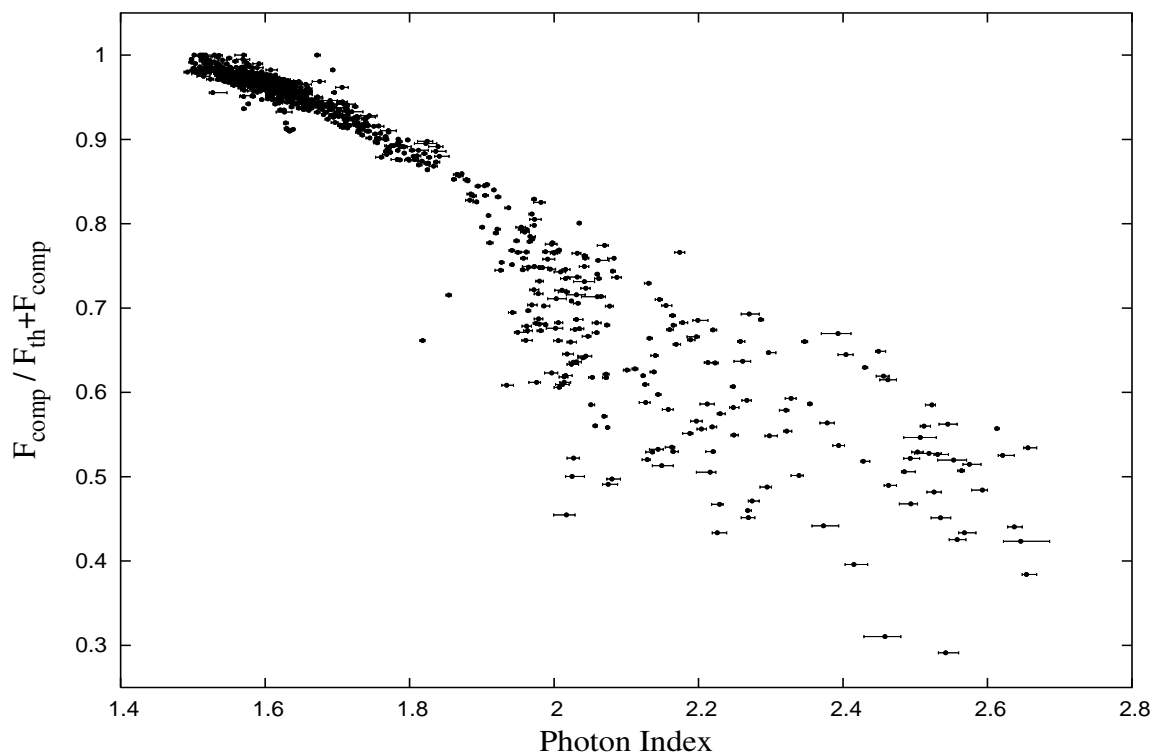}
\caption{
Observed ratio of the Comptonized photon flux  $F_{comp}$  and the total photon flux   $F_{comp}+F_{th}$  versus the photon index
$\Gamma$. In the low/hard state the soft photon (disk) radiation is fully Comptonized in the hot electron environment.
 The ratio steadily decreases toward the  soft state $\Gamma>2$. It is evident that the area of the Comptonized region (presumably converging
flow) shrinks as a source undergoes the state transition. 
}
\label{comp_ratio}
\end{figure}

\begin{figure}[ptbptbptb]
\includegraphics[width=4.in,height=6.1in,angle=-90]{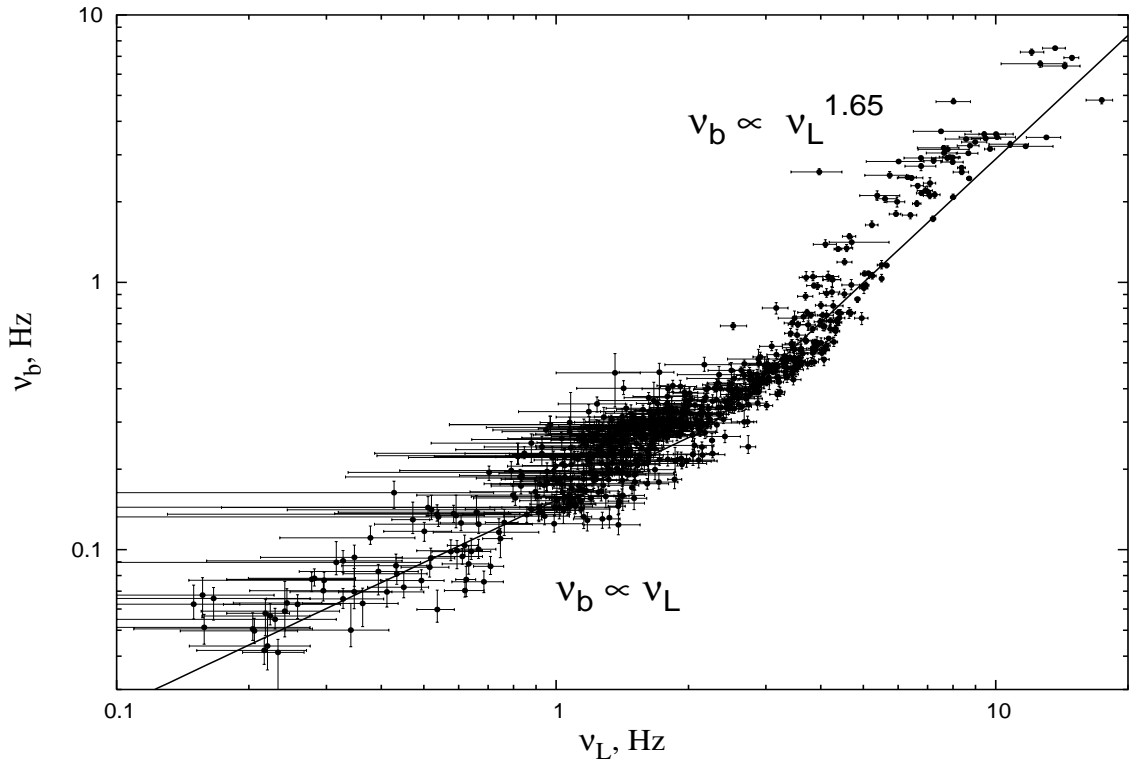}
\caption{
Observed correlation of break frequency $\nu_b$ vs low frequency $\nu_L$ which is fitted by the broken power law (solid line). 
For high frequency values ($\nu_L>2.2$ Hz) the power-law index is approximately 1.65 as for the low ones that is approximately 
1 (see text). 
}
\label{br_lowfr}
\end{figure}

\begin{figure}[ptbptbptb]
\includegraphics[scale=1.0,angle=0]{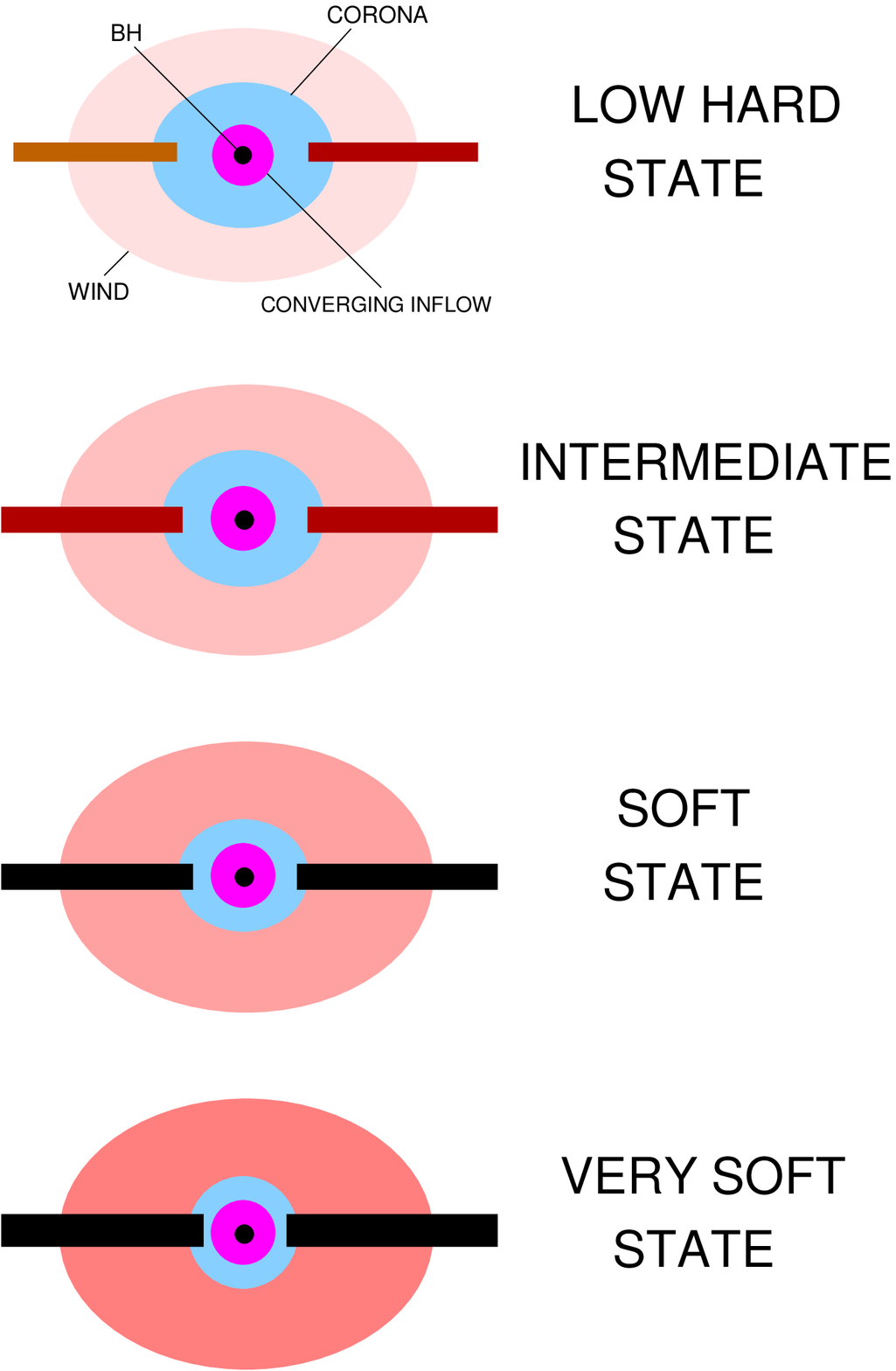}
\caption{
The inferred scenario of the spectral transition in Cyg X-1. Strength of disk and outflow (wind) increase towards the soft states. 
 }
\label{geometry}
\end{figure}

In Figure \ref{geometry} we present a scenario that is infered using our spectral and 
timing data analysis of Cyg X-1 spectral state transitions.
In the low/hard state X-ray radiation comes from a relatively extended area. A Compton cloud 
covers the large portion of the accretion disk that generates  soft photons.
Most of these photons are upscattered in the hot Compton cloud. On its way to the 
observer some fraction of the Comptonized radiation is reprocessed  
in the relatively transparent  cloud. Here  Fe K$_{\alpha}$ line is formed  
because the photons of energy  close to K shell ionization edge and higher are effectively
absorbed and then remitted (with a certain probability) as K$_{\alpha}$-photons.
 QPO features are not washed out by photon scattering in the wind shell in this state 
 because the optical depth of the wind is less then 0.5 (see Fig. \ref{line_index}, lower panel).
The wind becomes thicker in the intermediate and  soft states and it is  optically
thick in the very soft (thermal dominated) state. This picture is supported by the fact  that
the oscillation amplitude  and the QPO strength steadily decreases when the 
source proceeds to soft state and the QPO features are completely washed out 
in the very soft state (see Fig. \ref{power}). Also the strength of K$_{\alpha}$ (that is 
presumably generated in the wind) increases with the index (see Fig. \ref{line_index}, upper panel). 
On the other hand, the Compton cloud becomes very compact and cooler in softer states 
with respect to that in the low/hard state.  The Compton cloud size approaches that  of the converging 
inflow which is  a few Schwarzschild radii.  Also, in perfect agreement with our scenario, the
inferred fraction of the Comptonized component in the emergent spectrum (see Fig. \ref{comp_ratio}) 
decreases with the index (i.e. with mass accretion rate).  

There several other studies of the  X-ray spectral and timing properties of Cyg X-1 and other
 BHs which directly relate to the results of our paper.
Particularly, Pottschmidt et al. (2003, P03)  also found correlations between the photon index and QPO frequencies with a clear sign of the index saturation  at high
values of  the (low frequency) QPO(see Fig. 7, panel b in P03). We found that  our QPO frequencies 
$\nu_L$ and QPO frequencies $\nu_3$ in P03 are almost identical and
they show the similar correlations with the photon index $\Gamma$. P03's  frequencies
 $\nu_2$ are presumably very close to our break frequency $\nu_b$.
 
To obtain  a description of the X-ray photon spectrum, P03 like many other authors (see e.g. GCR99) 
 used an empirical model in which each source spectrum is a sum of power law spectrum with photon index
$\Gamma$ and a multi-temperature disk blackbody. To this continuum, a reflection spectrum after MZ95 was 
added. GCR99 emphasized that this empirical model is an oversimplification and therefore the best-fit 
parameter values do not necessarily represent  the physically meaningful quantities. 

GCR99 also showed that the reflection model can properly mimic the reprocessed 
component of the spectrum. Particularly for intermediate and soft states 
(when $R>1$ in the reflection modeling) one can clearly see the high strength in the
 fluorescent K$_{\alpha}$ iron line at 6-7 keV and the deep broad smeared iron
K-edge at $\sim 7-10$ keV (see Figs. 6-7 in GCR99).  We confirm GCR99's conclusions that 
these line and K-edge edge features are more pronounced towards soft states.
However,  we argue that these spectral features are results of reprocessing  of the central
 source hard radiation in the outflow warm absorber (wind) 
that surrounds the central BH. This our claim is supported by observed QPO power decay 
towards the softer states that is presumably a result of reprocessing of time signal in 
the extended relatively warm wind (see Fig \ref{power}).   

Recently Proga (2005) and Proga \& Kallman (2004) demonstrated using the results of hydro simulations, 
that the disk illumination by the radiation of the central source or just 
local disk luminosity can  launch a wind off the disk photosphere. They argue that radiation pressure due
 to UV (line resonances) couples the X-ray and UV radiation processes
 by driving the disk material above the
disk where the most part of the X-ray emission of the central source propagates.  
A strong disk wind develops when  the local  disk flux $L_D$ is more than 0.3 $L_{\rm Edd}$.  For a less luminous disk the line force can still force material off the disk
but it fails to accelerate the flow to escape velocity.    
 
\begin{figure}[ptbptbptb]
\includegraphics[scale=0.66,angle=-90]{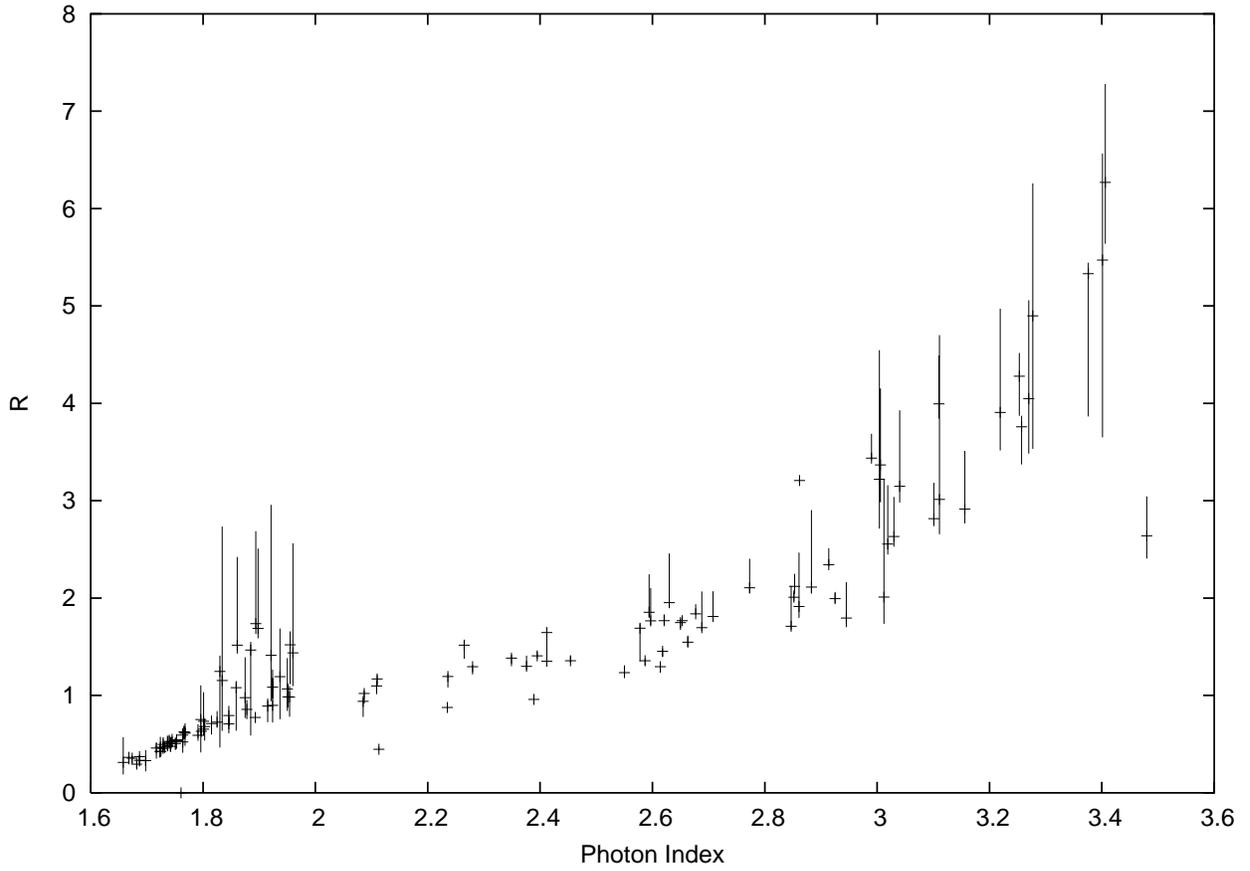}
\caption{
The inferred reflection scaling factor R as a function of the index using the MZ95 reflection model.
 }
\label{rel_refl}
\end{figure}

One should stress that the timing and spectral data do reveal signatures of reprocessed component 
of the hard radiation in terms of high iron line equivalent  width (about one keV in the soft states) and
 broad iron K-edge features.
 The strengths of these features seen in the observations strongly exclude their formation by
disk reflection.  The highest value of EW due to reflection must be  less than 100 eV
(see Basko 1978 and George \& Fabian 1991). 
In fact, the best-fit  values of the reflection scaling factor $R=1-5$ inferred from the Cyg X-1
 observations contradict to the basic assumption 
of the reflection model for which $R$ cannot be more than 0.25. 
As a result of our data analysis we have also found using the reflection model by MZ95 that $R$ correlates 
with spectral index
and values of $R$ are  as high as $6$ at $\Gamma\sim 3.4$ (see Fig \ref{rel_refl}).

In principle, the  reflection effect from outer relatively cold parts of the disk would be 
detected in the low state of the source when the spectrum is relatively hard
(photon index $\Gamma\sim 1.5$). In this case one can see a reflection hump in the 
spectrum around 10-15 keV (ST80). The main problem with the detection of this feature 
along with low  EW of the line is that the fraction of reflected emission $f$ in the 
resulting spectrum is about 8\% and less, i.e. $f=R\cdot A\lax0.25\cdot0.3=0.075$
where $A\lax 0.3$ is albedo of the cold material (Basko, Sunyaev \& Titarchuk 1974, Titarchuk 1987). 
Such  tiny reflection features will be readily washed out by intervening wind 
clouds (see above). The disk reflection effects become much weaker for softer spectra 
(states), for which photon index more than 2. 
Thus the strong iron and K-edge features detected in the soft state are likely not due 
to reflection from the disk at all. Rather they should be   formed in the extended wind clouds 
surrounding the central black hole. The relative large time lag observed in the soft state 
(see GCR99) really corroborate this scenario of reprocessing of the spectrum of the central object 
in the very extended environment (wind) of the source.

 McConnell et al. (2002)  using GRO/COMPTEL observations of Cyg X-1 claim that the high 
 energy tail  in the soft state shows no cutoff up to at least 0.5 MeV and  10 MeV respectively.  They  demonstrate that their analysis from BATSE, OSSE and COMPTEL show that the combined  spectrum
  can be described by a single power law with a best-fit photon index of $\Gamma=2.58\pm 0.03$.
 For these particular states for which $\Gamma>2.1$ {\it RXTE} power spectra are featureless 
 (see Fig. \ref{power}), no QPOs are seen in these states. In fact, Grove (2005, private communications) 
 confirms that OSSE power spectra of Cyg X-1 also do not show any signatures of QPO but only red noise.   It means that in these observations the central BH is not seen directly but  through a very 
  extended cloud that ultimately washes out  the timing and spectral information of the central object.
   It is also worth pointing out that the COMPTEL exposure time is about $6\times 10^6$ s that is much
   longer than the spectral formation time near the central black hole(which is  of order of crossing time
    of the innermost part of BH, a few times of $10^7 {\rm cm}/c\sim 10^{-3}$ s). 
 The hydrodynamical time scale can be longer by two orders of magnitude than the spectral 
 formation time  but it is still 7 orders of magnitude shorter than the COMPTEL exposure time. 
 One can not exclude that the high energy emission detected by   COMPTEL is probably not related to central BH, 
but it can be a result of some other process,
 for example in the outflow  or jets (see  the recent discovery of very high energy gamma rays 
 associated with an X-ray binary LS 5039 in a paper by Aharonian et al. 2005).  
 
\begin{figure}[ptbptbptb]
\includegraphics[scale=0.66,angle=-90]{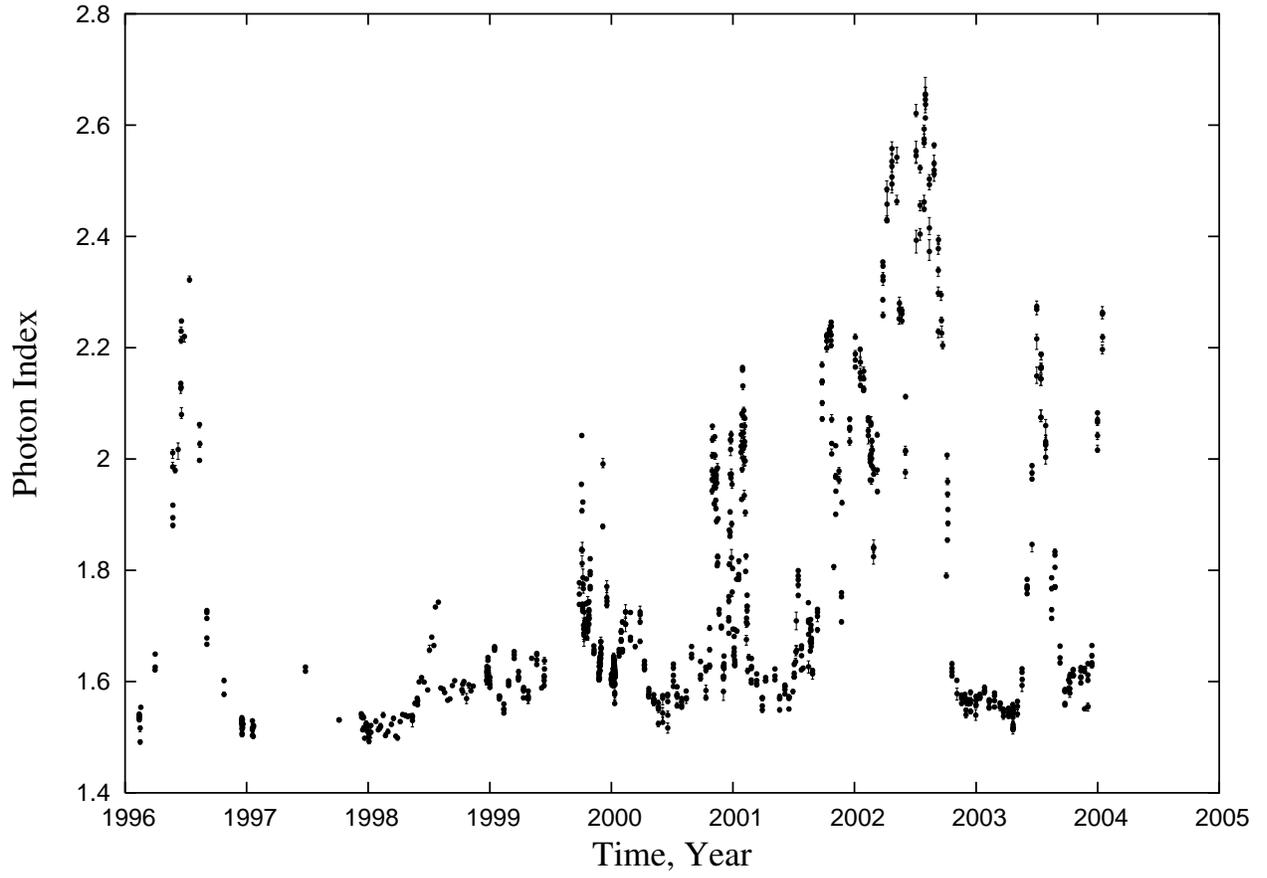}
\caption{
The variation of the photon index in Cyg X-1 throughout the entire RXTE mission.
 }
\label{index_vs_time}
\end{figure}

In Figure \ref{index_vs_time} we present   timescale photon index variation for different states
in  Cyg X-1 during $\sim$ 8 years of {\it RXTE} observations.  In particular, Cyg X-1 can be rather stable in the low/hard state for
the several months, while it is very dynamic in the soft and very soft states 
where it changes  its spectral index on timescale of a day. Thus  Cyg X-1 stays in 
the low/hard with an occasional transition (once per several years) to the  soft state where the power-law spectrum becomes significantly steeper 
(with $\Gamma>2$).  Also, one or two times per year   Cyg X-1 exhibits so called 
"failed state transions",  when it starts to transition but does not reach a soft state, stops at some
intermediate state  and falls back to a hard one.

We acknowledge productive discussions with Ralph Fiorito and Chris Shrader.
\newpage

\newpage

\newpage

\newpage

\newpage

\newpage

\newpage

\newpage

\newpage

\newpage

\newpage

\newpage

\newpage

\newpage

\newpage

\newpage

\newpage
\begin{deluxetable}{llllll}
\tablewidth{0pt}
\tablecaption{Summary of RXTE archive data on Cyg X-1}
\tablehead{\colhead{Proposal ID} & \colhead{Start Date} & \colhead{Stop Date} & \colhead{Time, sec}& \colhead{$N_{int}$}&\colhead{${\bar{N}_{PCUon}}$}}
\startdata
10235-01 & 12/02/1996 & 17/02/1996 & 9332.18 & 4 & 5.0  \\
10236-01 & 15/12/1996 & 18/12/1996 & 28239.78 & 13 & 5.0  \\
10412-01 & 22/05/1996 & 12/08/1996 & 18125.03 & 9 & 4.94  \\
10238-01 & 26/03/1996 & 03/02/1997 & 7475.19 & 3 & 3.65  \\
10240-01 & 12/02/1996 & 19/12/1996 & 44813.12 & 16 & 4.56  \\
10241-01 & 23/10/1996 & 24/10/1996 & 5088.0 & 2 & 4.37  \\
10257-01 & 08/06/1996 & 12/07/1996 & 2432.0 & 4 & 5.0  \\
10512-01 & 04/06/1996 & 18/06/1996 & 8012.22 & 6 & 5.00  \\
20173-01 & 17/01/1997 & 20/01/1997 & 26883.84 & 8 & 5.0  \\
20175-01 & 25/06/1997 & 02/01/1998 & 17783.69 & 7 & 4.99  \\
30155-01 & 22/12/1998 & 28/12/1998 & 38003.49 & 16 & 4.81  \\
30157-01 & 11/12/1997 & 09/12/1998 & 117347.25 & 47 & 4.85  \\
30158-01 & 10/12/1997 & 30/12/1997 & 30587.26 & 11 & 4.90  \\
30162-01 & 12/05/1998 & 10/10/1998 & 17892.08 & 7 & 5.00  \\
40099-01 & 14/01/1999 & 11/02/2000 & 115199.16 & 64 & 3.73  \\
40100-01 & 14/02/1998 & 02/14/2002 & 197798.51 & 90 & 3.82  \\
40101-01 & 27/09/1999 & 10/10/1999 & 23505.19 & 19 & 3.93  \\
40102-01 & 05/01/2000 & 10/01/2000 & 234309.48 & 84 & 3.09  \\
40417-01 & 25/04/1999 & 13/06/1999 & 11004.28 & 8 & 2.92  \\
50110-01 & 11/02/2000 & 06/04/2002 & 387571.88 & 210 & 3.22  \\
50109-01 & 22/12/2000 & 15/02/2001 & 59020.71 & 40 & 3.90  \\
50109-03 & 05/11/2000 & 10/01/2001 & 60428.65 & 22 & 3.64  \\
50119-01 & 28/10/2000 & 04/01/2001 & 48984.88 & 23 & 3.82  \\
60089-01 & 19/08/2001 & 30/08/2001 & 31545.51 & 13 & 2.80  \\
60089-02 & 23/09/2001 & 28/10/2001 & 14409.54 & 6 & 3.11  \\
60089-03 & 07/10/2001 & 21/02/2002 & 60380.83 & 24 & 3.15  \\
60090-01 & 08/03/2002 & 03/04/2004 & 363880.25 & 189 & 2.88  \\
60091-01 & 15/10/2001 & 22/10/2001 & 26984.65 & 12 & 3.85 \\
60136-03 & 28/06/2001 & 09/07/2001 & 2512.0 & 5 & 3.12  \\
70015-04 & 15/09/2002 & 17/09/2002 & 7685.78 & 3 & 3.68  \\
70414-01 & 30/07/2002 & 29/12/2002 & 21393.17 & 14 & 4.12  \\
80111-01 & 19/04/2003 & 20/04/2003 & 38834.50 & 13 & 3.30  \\

\enddata
\end{deluxetable}

\newpage

\begin{deluxetable}{l}
\tablewidth{0pt}
\tabletypesize{\footnotesize}
\tablecaption{Results of fitting PCA/HEXTE spectrum with BMC model.}
\tablehead{ \colhead{MODEL/PARAMETER   \phantom{aaaaaaaaaaaaaaaaa}                            VALUE }}
\startdata
\phantom{aaaaaaaaaaaaaaaa}\large Low/Hard State\tablenotemark{a}  \\
\hline
BMC\\
\phantom{aaaa}$\Gamma$,\phantom{aaaaaaaaaaaaaaaaaa.aaaaaaaaaaaaa}   $1.47\pm0.01$ \\
\phantom{aaaa}$kT_{col}$, keV \phantom{.aaaaaaaaaaaaaaaaaaaaaaa}  $0.73_{-0.06}^{+0.04}$\\
\phantom{aaaa}A \phantom{aaaaaaaaaaaaaaaaaaaaaaaaaaaaaaa} $2.04\pm0.06$\\
\hline
GAUSSIAN (6.4 keV fixed)\\
\phantom{aaaa}$\sigma$, \phantom{aaaaaaaaaaaaaaaaaaaaaaaaaaaaaaa} $0.71\pm0.17$ \\
\phantom{aaaa}EW, eV\phantom{aaaaaaaaaaaaaaaaaaaaaaaaaa} $198 $ \\
\hline
HIGHECUT\\
\phantom{aaaa}$E_{cut}$, keV \phantom{aaaaaaaaaaaaaaaaaaaaaaaa} $33.8_{-2.6}^{+3.6}$\\
\phantom{aaaa}$E_{fold}$, keV \phantom{aaaaaaaaaaaaaaaaaaaaaaa} $212\pm13$\\
\hline
CONSTANT (Cross-Normalization)\\
\phantom{aaaa}PCA/HEXTE A  \phantom{aaaaaaaaaaaaaaaaaa} $0.81\pm0.01$\\
\phantom{aaaa}PCA/HEXTE B  \phantom{aaaaaaaaaaaaaaaaaa} $0.81\pm0.01$ \\
\hline
\phantom{aaaa}$\chi^2_\nu\, (N_{dof})$ \phantom{aaaaaaaaaaaaaaa.aaaaaaaaa}  0.84 (159)\\
\hline
  \\
\phantom{aaaaaaaaaaaaaaaa}\large Intermediate State\tablenotemark{b}  \\
\hline
BMC\\
\phantom{aaaa}$\Gamma$,\phantom{aaaaaaaaaaaaaaaaaa.aaaaaaaaaaaaa}   $1.83\pm0.01$ \\
\phantom{aaaa}$kT_{col}$, keV \phantom{.aaaaaaaaaaaaaaaaaaaaaaa}  $0.56\pm0.02$\\
\phantom{aaaa}A \phantom{aaaaaaaaaaaaaaaaaaaaaaaaaaaaaaa} $1.08\pm02$\\
\hline
GAUSSIAN (6.4 keV fixed)\\
\phantom{aaaa}$\sigma$, \phantom{aaaaaaaaaaaaaaaaaaaaaaaaaaaaaaa} $1.46\pm0.09$ \\
\phantom{aaaa}EW, eV\phantom{aaaaaaaaaaaaaaaaaaaaaaaaaa} $930 $ \\
\hline
HIGHECUT\\
\phantom{aaaa}$E_{cut}$, keV \phantom{aaaaaaaaaaaaaaaaaaaaaaaa} $31.2_{-1.8}^{+3.1}$\\
\phantom{aaaa}$E_{fold}$, keV \phantom{aaaaaaaaaaaaaaaaaaaaaaa} $181\pm12$\\
\hline
CONSTANT (Cross-Normalization)\\
\phantom{aaaa}PCA/HEXTE A  \phantom{aaaaaaaaaaaaaaaaaa} $0.86\pm0.01$\\
\phantom{aaaa}PCA/HEXTE B  \phantom{aaaaaaaaaaaaaaaaaa} $0.85\pm0.01$ \\
\hline
\phantom{aaaa}$\chi^2_\nu\, (N_{dof})$ \phantom{aaaaaaaaaaaaaaa.aaaaaaaaa}  0.90 (159)\\
\hline
   \\
\phantom{aaaaaaaaaaaaaaaa}\large Very Soft State \tablenotemark{c}\\
\hline
BMC\\
\phantom{aaaa}$\Gamma$,\phantom{aaaaaaaaaaaaaaaaaa.aaaaaaaaaaaaa}   $2.64\pm0.02$ \\
\phantom{aaaa}$kT_{col}$, keV \phantom{.aaaaaaaaaaaaaaaaaaaaaaa}  $0.51\pm0.02$\\
\phantom{aaaa}A \phantom{aaaaaaaaaaaaaaaaaaaaaaaaaaaaaaa} $0.66\pm0.02$\\
\hline
GAUSSIAN (6.4 keV fixed)\\
\phantom{aaaa}$\sigma$, \phantom{aaaaaaaaaaaaaaaaaaaaaaaaaaaaaaa} $1.33\pm0.06$ \\
\phantom{aaaa}EW, eV\phantom{aaaaaaaaaaaaaaaaaaaaaaaaaa} $1366 $ \\
\hline
HIGHECUT\\
\phantom{aaaa}$E_{cut}$, keV \phantom{aaaaaaaaaaaaaaaaaaaaaaaa} $26.7_{-5.1}^{+6.0}$\\
\phantom{aaaa}$E_{fold}$, keV \phantom{aaaaaaaaaaaaaaaaaaaaaaa} $149^{+138}_{-51}$\\
\hline
CONSTANT (Cross-Normalization)\\
\phantom{aaaa}PCA/HEXTE A  \phantom{aaaaaaaaaaaaaaaaaa} $0.91\pm0.03$\\
\phantom{aaaa}PCA/HEXTE B  \phantom{aaaaaaaaaaaaaaaaaa} $0.88\pm0.03$ \\
\hline
\phantom{aaaa}$\chi^2_\nu\, (N_{dof})$ \phantom{aaaaaaaaaaaaaaa.aaaaaaaaa}  1.16 (159)\\
\enddata
\tablenotetext{a}{ObsID:30158-01-03-00 (Dec 14, 1997)}
\tablenotetext{b}{ObsID:50119-01-04-01 (Dec 19, 2000)}
\tablenotetext{c}{ObsID:60090-01-11-01 (Jul 26, 2002)}
\label{statestab}
\end{deluxetable}

\end{document}